\documentclass[twocolumn]{aastex631}
\usepackage{color}
\usepackage[titletoc]{appendix}
\usepackage{amsmath}
\usepackage{amssymb}
\usepackage{mathtools}
\usepackage{upgreek}
\usepackage{float}
\usepackage{comment}
\usepackage{enumitem}
\usepackage{natbib}
\usepackage{graphicx}
\usepackage{bm}
\usepackage{totcount}
\usepackage{multirow}

\newtotcounter{citnum} 
\def\oldbibitem{} \let\oldbibitem=\bibitem
\def\bibitem{\stepcounter{citnum}\oldbibitem}
\defcitealias{2018AJ....155..165P}{P18}

\shortauthors{Millholland et al.}
\shorttitle{Constraining $Q_{\star}'$}

\begin{document} 

\title{Empirical Constraints on Tidal Dissipation in Exoplanet Host Stars}

\author[0000-0003-3130-2282]{Sarah C. Millholland}
\affiliation{Department of Physics, Massachusetts Institute of Technology, Cambridge, MA 02139, USA}
\affiliation{MIT Kavli Institute for Astrophysics and Space Research, Massachusetts Institute of Technology, Cambridge, MA 02139, USA}
\email{sarah.millholland@mit.edu}

\author[0000-0002-1417-8024]{Morgan MacLeod}
\affiliation{Center for Astrophysics, Harvard \& Smithsonian, 60 Garden Street, MS-16, Cambridge, MA 02138, USA}

\author{Felicia Xiao}
\affiliation{Department of Physics, Massachusetts Institute of Technology, Cambridge, MA 02139, USA}

\begin{abstract}
The orbits of short-period exoplanets are sculpted by tidal dissipation.  However, the mechanisms and associated efficiencies of these tidal interactions are poorly constrained. We present robust constraints on the tidal quality factors of short-period exoplanetary host stars through the usage of a novel empirical technique. The method is based on analyzing structures in the population-level distribution of tidal decay times, defined as the time remaining before a planet spirals into its host star due to stellar tides. Using simple synthetic planet population simulations and analytic theory, we show that there exists a steady-state portion of the decay time distribution with an approximately power-law form. This steady-state feature is clearly evident in the decay time distribution of the observed short-period planet population. We use this to constrain both the magnitude and frequency dependence of the stellar tidal quality factor and show that it must decrease sharply with planetary orbital period. Specifically, with $Q_{\star}' = Q_0 (P/2 \ \mathrm{days})^{\alpha}$, we find $10^{5.5} \lesssim Q_0 \lesssim 10^7$ and $-4.33 \lesssim \alpha \lesssim -2$. Our results are most consistent with predictions from tidal resonance locking, in which the planets are locked into resonance with a tidally excited gravity mode in their host stars.
\end{abstract}

\section{Introduction}
\label{sec: Introduction}

Short-period exoplanets exist in a regime where tidal forces can generate drastic long-term orbital, spin, and even physical evolution. When a planet is in close proximity with its host star, both bodies experience a tidal deformation. A non-zero eccentricity and/or spin-axis tilt (``obliquity'') of the planet causes its tidal deformation to vary in time, producing tidal heating in the planetary interior. These \textit{planetary tides} generally dominate the transfer of energy and angular momentum and cause the planetary orbit to circularize and decay inwards. If a planet then reaches a damped state of zero eccentricity and obliquity, further evolution will still occur but as a result of \textit{stellar tides}, or the tidal deformation in the star raised by the planet. A mismatch between the stellar rotation period $P_{\mathrm{rot}}$ and the planetary orbital period $P$ causes tidal dissipation inside the star. If $P_{\mathrm{rot}} > P$, as is usually the case in main-sequence or evolved planetary systems, the tidal torques cause angular momentum transfer from the orbit to the star, decreasing $P$ and increasing $P_{\mathrm{rot}}$. The end state of this process is for the planet to eventually spiral in and be engulfed by its star. This is the regime that many close-in exoplanets, particularly hot Jupiters, are found in.

There has been a vast accumulation of observational evidence that both planetary tides and stellar tides are at work in close-in exoplanetary systems. Most directly, the orbital decay of hot Jupiters WASP-12 b \citep{2016A&A...588L...6M, 2017AJ....154....4P, 2020ApJ...888L...5Y} and Kepler-1658 b \citep{2022ApJ...941L..31V} have been measured in real time via transit time shifts, indicating they will be engulfed by their stars in only a few million years. \cite{2023Natur.617...55D} reported an infrared transient that appears to indicate the exact moment of a planet's engulfment by a star, an event which is likely to be the end state of tidal inspiral. 

Ample, albeit less direct, evidence of tidal evolution has come through population-level trends. We will focus first on the effects of stellar tides. \cite{2009ApJ...698.1357J} identified features in the semi-major axis and age distribution of a planet sample (predominantly hot Jupiters) that indicated tidal sculpting; older planets tended to be farther from their host stars. \cite{2014ApJ...786..139T} found a dearth of planets with short periods around rapidly rotating stars, and \cite{2024arXiv240207893C} found the same around evolved stars. As the related precursor to engulfment, there is evidence that the stars hosting some of the shortest-period planets have been tidally spun up \citep[e.g.][]{2009MNRAS.396.1789P, 2012MNRAS.422.3151H, 2015A&A...577A..90M, 2018AJ....155..165P}. Furthermore, by comparing the Galactic velocity dispersion of hot Jupiter host stars with a matched sample of field stars, \cite{2019AJ....158..190H} showed that the hot Jupiter systems tend to be younger, which is consistent with the picture that hot Jupiters are destroyed by tides while their stars are still on the main sequence. \cite{2022A&A...658A.199M} and \cite{2024AJ....168....7B} obtained similar findings using kinematics-based analyses. \cite{2023PNAS..12004179C} and \cite{2023AJ....166..209M} recently found a consistent result when they measured the occurrence rate of hot Jupiters as a function of age, stellar mass, and metallicity and found evidence that the occurrence rate decreases with stellar age.

Despite this evidence that tidal evolution is sculpting close-in planets, there is still significant uncertainty about the mechanisms and associated efficiencies of tidal dissipation. The detailed processes by which tidal energy is dissipated is infamously challenging to understand from first principles, and many theoretical mechanisms have been proposed. These can be broadly categorized into two major components: equilibrium tides and dynamical tides \citep[e.g.][]{2008EAS....29...67Z}. Equilibrium tides refer to the quasi-hydrostatic deformation response that is thought to be dissipated by turbulent viscosity in convective zones. Dynamical tides are oscillatory modes excited by the periodic tidal potential, and they come in different flavors depending on whether they propagate in convective or radiative zones \citep[e.g.][]{2014ARA&A..52..171O}. The efficiency of tidal dissipation is typically parameterized as the dimensionless reduced tidal quality factor $Q' = 3 Q/(2 k_2)$, where $Q$ is proportional to the ratio between the maximum energy stored in the tide and the energy dissipated per period \citep[e.g.][]{1966Icar....5..375G} and $k_2$ is the tidal Love number. $Q'$ is an inverse measure of dissipation, so larger $Q'$ indicates less efficient dissipation. There is not yet consensus on the overall magnitude and tidal frequency dependence of the stellar $Q_{\star}'$ of planet-hosting stars. Crucially, constraints on $Q_{\star}'$ would have important implications for deciphering tidal mechanisms. 

In this work, we introduce a novel empirical method as a means of constraining the $Q_{\star}'$ of stars hosting short-period planets, mostly hot Jupiters. The method relies on population-level features that appear when studying the distribution of timescales of tidally-induced orbital decay. It allows us to constrain the overall magnitude and frequency dependence of the stellar tidal quality factors, which offers insight onto the dominant mechanisms of tidal dissipation. Moreover, it also allows for a concise and concrete means of summarizing the effects of tides at the population level and making predictions of expected rates of tidal orbital decay and engulfment. Our work is similar in spirit to \cite{2010ApJ...723..285H, 2012ApJ...757....6H}, who used the sample of observed exoplanets to calibrate a model for both stellar tides and planetary tides based on the planet mass, eccentricity, and semi-major axis distribution. We revisit similar tidal models with an expanded exoplanet population but use a different calibration that allows us to probe deeply into the phase space of tidally-decaying orbits. 

For a point of clarity, we note that this work pertains exclusively to stellar tides. As for planetary tides, there is an extensive literature on their effects on exoplanetary orbital, spin, and interior evolution. Examples include the formation of hot Jupiters and other planets through high-eccentricity tidal migration \citep[e.g.][]{1996Sci...274..954R, 2007ApJ...669.1298F, 2011ApJ...735..109W, 2018ARA&A..56..175D}, the eccentricity distribution of close-in planets \citep[e.g.][]{2008ApJ...686L..29M, 2012MNRAS.425..757K, 2012MNRAS.422.3151H, 2018MNRAS.477..175O, 2023MNRAS.525..876M, 2024MNRAS.527.8245L}, formation of ultra-short period planets \citep{2019AJ....157..180P, 2019MNRAS.488.3568P, 2020ApJ...905...71M}, repulsion of planets near mean-motion resonances \citep{2012ApJ...756L..11L, 2013AJ....145....1B, 2019NatAs...3..424M}, and tidal locking \citep{2017CeMDA.129..509B}, to name a few. In general, short-period exoplanets are sculpted by both planetary tides and stellar tides \citep[e.g.][]{2010ApJ...723..285H}, but planetary tides generally dominate the planet's semi-major axis, eccentricity, and spin evolution when the orbit is eccentric and/or the planet's spin axis is tilted. Thus, none of the examples described above offer significant constraints on the dissipation efficiency in the host star. This paper focuses on the cases where the eccentricity and obliquity have damped to zero and stellar tides are all that remains. This regime is relevant to the very shortest-period planets, particularly hot Jupiters, which usually have circular orbits. 

Our analysis proceeds as follows. We first use experiments with simulated hot Jupiter populations to illustrate distinct, tidally-sculpted features that emerge at the population level (Section \ref{sec: toy model}). We show that this behavior is summarized succinctly by an analytic model involving a continuity equation (Section \ref{sec: analytic model}). In Section \ref{sec: generalization}, we introduce two approaches to parameterizing $Q_{\star}'$. We employ these in Section \ref{sec: observations} on observed planet samples and use the population-level features to obtain empirical constraints on the magnitude and frequency dependence of $Q_{\star}'$. These constraints are then compared to theoretical models of tidal dissipation (Section \ref{sec: discussion}). Code reproducing some of our results can be found at \url{https://github.com/smillholland/Hot_Jupiter_Tides}.

\newpage
\section{Simulated Hot Jupiter Populations}
\label{sec: toy model}

We begin by introducing a toy model based on simulated hot Jupiter populations, with the goal of exploring population-level features of tidal decay. We will build the complexity of the simulations gradually, in an effort to clearly illustrate the dependence of our results on the various assumptions. Our first experiment is the simplest possible mock population of hot Jupiter systems, featuring uniform physical parameters with $M_{\star} = M_{\odot}$, $R_{\star} = R_{\odot}$, and $M_p = M_{\mathrm{Jup}}$. The semi-major axes are assigned randomly from $0.01-0.1$ AU. The planets are also assigned ages randomly from $0-8$ Gyr, so the total number of planets in the population rises over time.

Assuming the planets' orbits circularize quickly and that they have zero planetary axial tilt, the only source of tidal dissipation is within the stars due to the tides raised by the planets.\footnote{We can also make this more explicit by defining the initial semi-major axes noted above as those that the planets attain as soon as they arrive at circular orbits. Thus, if many hot Jupiters form via high-eccentricity migration, this would be the semi-major axis that they obtain right after that process has ended.} We model the effects of stellar tides using a traditional equilibrium tides framework with constant $Q_{\star}'$. (More complex frameworks are explored later in the paper.) The instantaneous orbital decay rate is given by \citep{1966Icar....5..375G, 2009ApJ...698.1357J, 2010A&A...516A..64L}
\begin{equation}
\begin{split}
\frac{1}{\tau_a} \equiv \frac{\dot{a}}{a} &= -\frac{9}{2}n\left(\frac{M_p}{M_{\star}}\right)\left(\frac{R_{\star}}{a}\right)^5\frac{1}{Q_{\star}'} \\
&= -\frac{9}{2}\left(\frac{G}{M_{\star}}\right)^{1/2}\frac{R_{\star}^5 M_p}{Q_{\star}'} a^{-13/2}.
\label{eq: tau_a}
\end{split}
\end{equation}
Here $G$ is the gravitational constant, $M_{\star}$ is the stellar mass, $R_{\star}$ is the stellar radius, $Q_{\star}' = 3 Q_{\star}/2 k_2$ is the reduced stellar tidal quality factor, $M_p$ is the planet mass, and $a$ is the semi-major axis. This equation assumes that the planet's orbital period is much shorter than the star's rotation period. Written differently, we have
\begin{equation}
\begin{split}
\frac{1}{\tau_a}  &= - \frac{1}{2.162 \ \mathrm{Gyr}}\left(\frac{M_{\star}}{M_{\odot}}\right)^{-\frac{1}{2}}\left(\frac{R_{\star}}{R_{\odot}}\right)^5\left(\frac{M_p}{M_{\mathrm{Jup}}}\right) \\
& \times \left(\frac{a}{0.03\mathrm{AU}}\right)^{-13/2}\left(\frac{Q_{\star}'}{10^6}\right)^{-1}.
\label{eq: tau_a unit version}
\end{split}
\end{equation}
In our starting mock population, we assign $Q_{\star}' = 10^6$ for all stars and assume that $Q_{\star}'$ does not change as a function of frequency. We note that the parameterization of the equilibrium tide relies on bulk, directly-observable stellar quantities like $M_\star$ and $R_\star$, and focuses all properties of the dissipation in $Q_\star'$. Other models incorporate predictions of theoretical stellar interior models for stars of differing mass, for example the mass of the convective zone and the convective velocity of its motions \citep[e.g.][]{2012ApJ...757....6H}.

The timescale $\tau_a$ is the \textit{instantaneous} decay rate when the planet is at a given semi-major axis. It is more physically meaningful to work with the timescale that it takes for the planet to move from its current orbit to disruption. We call this simply the ``decay timescale'', which is given by
\begin{equation}
\tau_d = -\frac{2}{13}\tau_{a}
\label{eq: tau_decay}
\end{equation}
assuming (for now) that $Q_{\star}'$ has no dependence on the planet's period. Note that $\tau_d$ is a positive quantity and $\tau_a$ is negative when $P_{\mathrm{rot}} > P$. Planets are are ``born'' with a particular $\tau_a(t=0)$ and $\tau_d(t=0)$, but these quantities evolve in time as the planet's semi-major axis decays.

\begin{figure}
\centering
\includegraphics[width=\columnwidth]{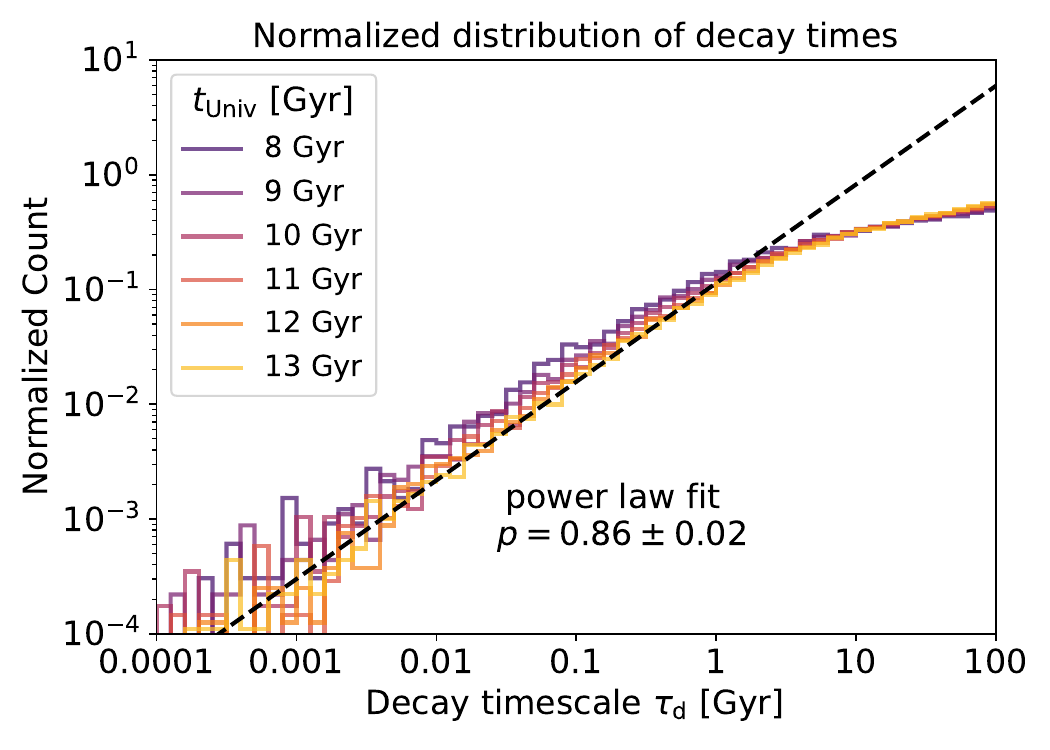}
\includegraphics[width=\columnwidth]{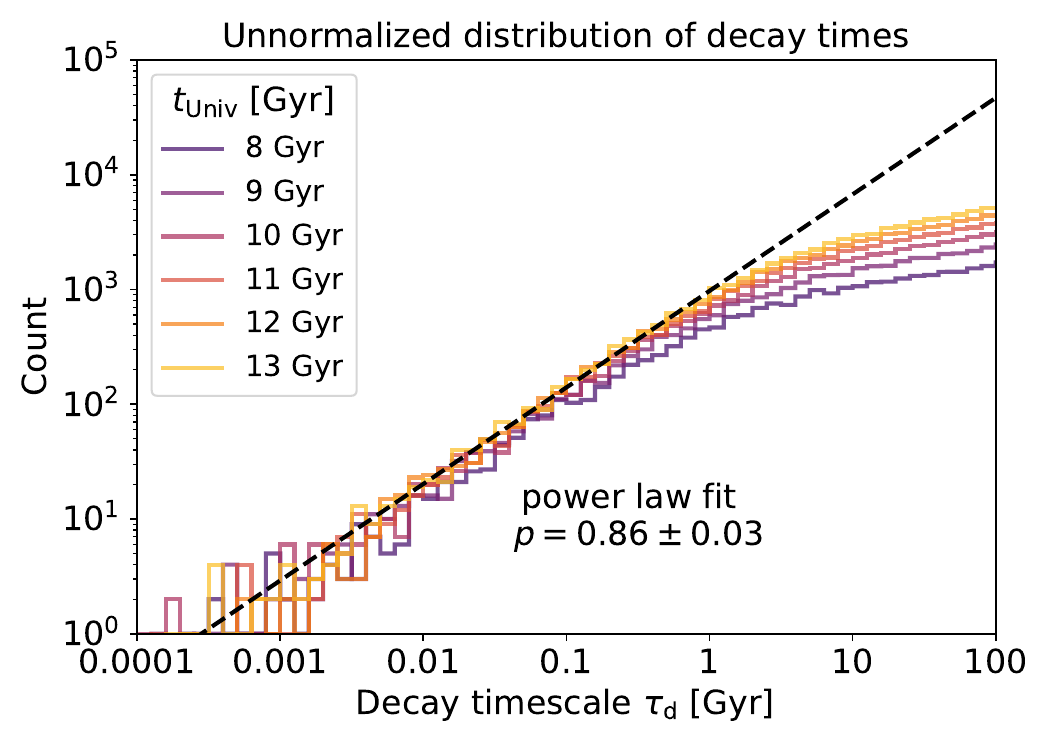}
\caption{Distributions of $\tau_d$ at a variety of time epochs of the Universe, $t_{\mathrm{Univ}}$, ranging from 8 Gyr to 13 Gyr. The dashed black line indicates a power law fit within the region $0.001 \ \mathrm{Gyr} < \tau_d < 1 \ \ \mathrm{Gyr}$, and the figure shows the mean and standard deviation of the power law exponent over 10 simulations. The version in the top panel has been normalized by the count, while the version in the bottom panel is unnormalized.} 
\label{fig: tau_decay distribution}
\end{figure}

Figure \ref{fig: tau_decay distribution} shows the evolution of the distribution of $\tau_d$ for the mock population of hot Jupiters. We show both the normalized and unnormalized distributions at a variety of time epochs of the Universe, ranging from 8 Gyr to 13 Gyr. Remarkably, both the normalized and unnormalized distributions of $\tau_d$ remain approximately in a steady state for $\tau_d \lesssim 1$ Gyr. The unnormalized distribution is a bit more constant than the normalized distribution in that region. The steadiness of the distribution in the $\tau_d \lesssim 1$ Gyr region is important since it indicates that stellar tides act in such a way that preserves the overall shape of the distribution over time. Even though individual planets in the distribution are born, move to lower $\tau_d$ as they age, and (given enough time) exit the distribution at the low $\tau_d$ end, the shape is preserved at the population level.

Figure \ref{fig: tau_decay distribution} shows that, for $\tau_d \lesssim 1$ Gyr, the curves of counts vs. $\tau_d$ are nearly linear in log-log space. We fit the region $0.001 \ \mathrm{Gyr} < \tau_d < 1 \ \ \mathrm{Gyr}$ using a power law and find that both the normalized and unnormalized distributions are well-fit by 
\begin{equation}
f \propto  {\tau_d}^p,
\label{eq: f(log10 tau)}
\end{equation}
where $f\Delta(\log_{10}\tau_d)$ represents the number of planets in bins of $\log_{10}\tau_d$, and $p$ is the power law exponent. In this paper we will refer to the $\tau_d \lesssim 1$ Gyr region as the ``steady-state region'' and $p$ as the ``slope of the steady-state region'', since it represents the slope of a linear fit to the distribution in log-log space. Using 10 trials and averaging over them, we find $p=0.86 \pm 0.02$ for the normalized distributions and $p=0.86 \pm 0.03$ for the unnormalized distributions.

\begin{figure}
\centering
\includegraphics[width=\columnwidth]{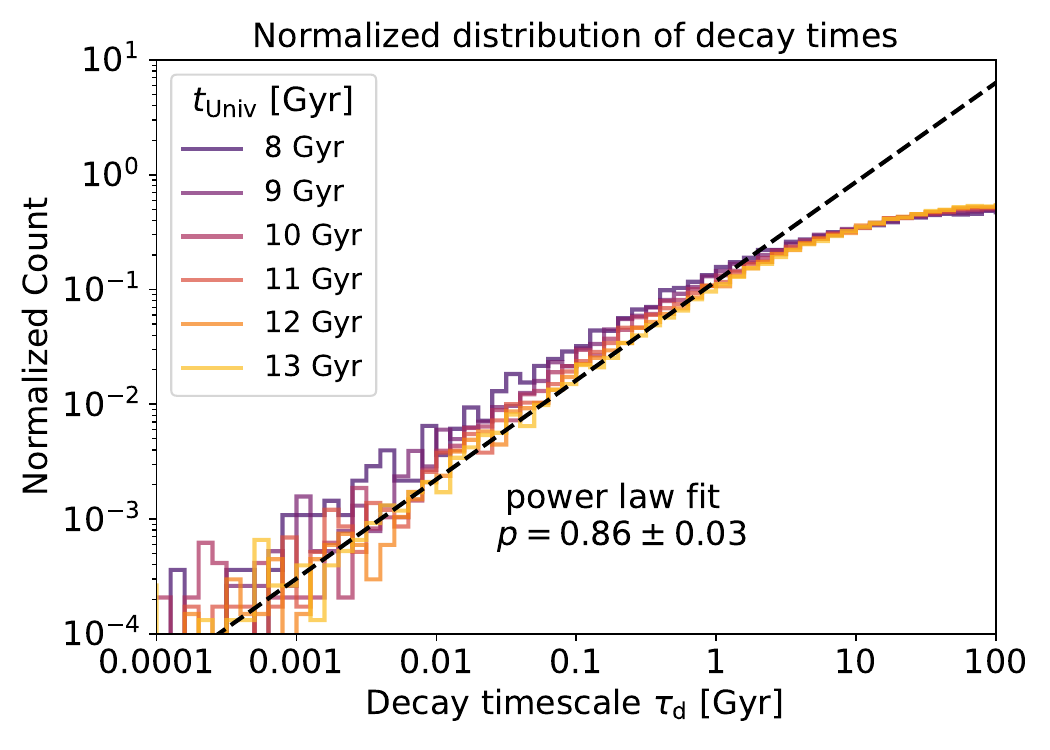}
\includegraphics[width=\columnwidth]{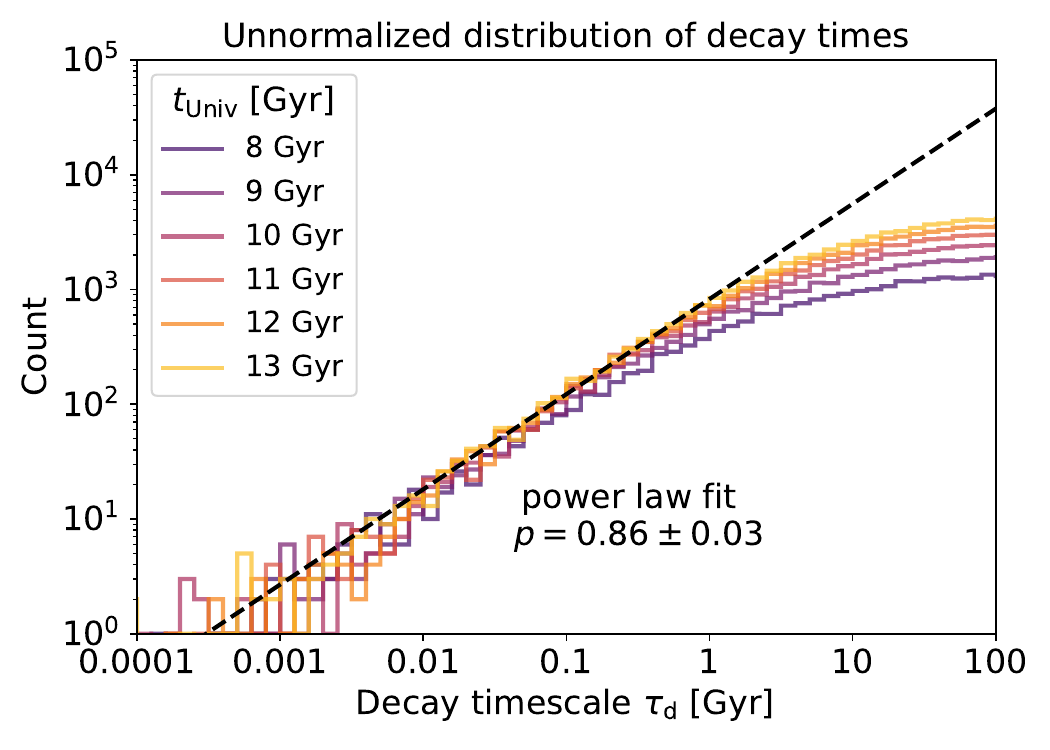}
\caption{Same as Figure \ref{fig: tau_decay distribution} except with randomized stellar and planetary parameters, as described in the text. The steady-state region is virtually identical to the earlier case, despite using randomized parameters. } 
\label{fig: tau_decay distribution with randomized params}
\end{figure}

The simulation just described is highly simplified given its usage of a fixed set of stellar and planetary parameters. We can dial up the complexity to examine the sensitivity of the $\tau_d$ distribution to the parameter choices. We explore randomized stellar and planetary parameters as follows: $M_{\star}/M_{\odot} \sim \mathcal{U}[0.7, 2.0]$ (where here and elsewhere $\mathcal{U}$ denotes a uniform distribution), $R_{\star}/R_{\odot} \sim \mathcal{U}[0.7, 2.0]$, $\log_{10}Q_{\star}' \sim \mathcal{U}[6, 8]$,  $M_p/M_{\mathrm{Jup}} \sim \mathcal{U}[0.5, 2.0]$, $a/\mathrm{AU} \sim \mathcal{U}[0.01, 0.1]$, and age/Gyr $\sim \mathcal{U}[0,8]$. (The $a$ and age distributions are the same as before). Figure \ref{fig: tau_decay distribution with randomized params} shows the $\tau_d$ distribution for the case of randomized parameters. Even with the randomized parameters, the distributions appear very similar to those from the earlier simulation. The slope of the steady-state region is the same as that found earlier. 

If we instead keep all of the parameters the same but randomize the semi-major axis according to a normal distribution, $a/{\mathrm{AU}}\sim\mathcal{N}(\mu = 0.04, \sigma = 0.015)$, then the distribution does change slightly. The slope of the steady-state region becomes $p = 0.93 \pm 0.02$. We explored further variations of the sampling scheme, which are reported in Appendix \ref{sec: source distribution}. We find that the slope of the steady-state region depends on the choice of the initial semi-major axis distribution, but the differences are not significant enough to affect the conclusions in this work. The reasoning behind this will become clearer in Section \ref{sec: analytic model} when we develop an analytic model. \\

\subsection{Disruption rate}
\label{sec: disruption rate}

The power law distribution $f \propto {\tau_d}^p$ in the $\tau_d \lesssim 1$ Gyr steady-state region can be used to find the rate at which planets are disrupted by falling into their host stars. Within a length of time $\Delta t$, the planets that are disrupted are those with $\tau_d \lesssim \Delta t$. Using equation \ref{eq: f(log10 tau)}, the number of disrupted planets is equal to 
\begin{equation}
N_{\mathrm{dis}} = \frac{A}{p\log 10}(\Delta t)^p,
\end{equation}
where we label $A$ as the constant of proportionality in equation \ref{eq: f(log10 tau)}.
We can then take the ratio between $N_{\mathrm{dis}}$ and the number of planets with $\tau_d$ below some threshold, $\tau_d \lesssim \tau_{d, \max}$, which we will call $N_{<\tau_{d,\max}}$. The ratio is equal to 
\begin{equation}
\frac{N_{\mathrm{dis}}}{N_{<\tau_{d,\max}}} = \left(\frac{\Delta t}{\tau_{d,\max}}\right)^p.
\label{eq: disruption rate}
\end{equation} 
Note that the constant $A$ has dropped out of the ratio, which now just depends on the power law exponent $p$. This result provides a simple and powerful means of estimating the number of disrupted planets in some length of time based on the number of planets in the population with $\tau_d < \tau_{d,\max}$. We need only identify a subset of the distribution that obeys the power law form $f \propto {\tau_d}^p$.  

\section{Analytic Model}
\label{sec: analytic model}

Thus far, we have characterized the distribution of decay times using simulated planet populations. Here we take a more fundamental approach and study the $\tau_d$ distribution and its steady-state region analytically. Although developed independently, our approach bears some resemblance to that of \cite{2012ApJ...750..106S}, who considered the problem of high-eccentricity tidal migration of giant planets and argued that the distribution of migrating planets with high eccentricities should be in an approximate steady state. The physical set-up is different in our problem, but in both cases, a continuity equation can be used to describe the planet population. 

Let $\rho(x, t)$ represent the number density of planets as a function of $x$ and time $t$ such that $dN = \rho(x, t) dx$ is the number of planets within the range $x$ to $x+dx$ at time $t$. The variable $x$ is itself a function of $\tau_d$, and we will here consider two variations: $x = \tau_d$ and $x = \ln\tau_d$. We will show that the two formulations lead to equivalent results. The continuity equation governs the evolution of $\rho(x,t)$ as planets are born into the population and ``flow'' from higher to lower $\tau_d$ over time. This yields
\begin{equation}
\frac{\partial\rho}{\partial t} + \frac{\partial(\rho u)}{\partial x} = S(x),
\label{eq: continuity equation}
\end{equation}
where $u \equiv dx/dt$ and $S(x)$ is the ``source distribution'', which describes the number of planets born per unit $x$ per unit time. The source distribution accounts for the initial semi-major axis distribution. 

Let us first consider the case where $x = \tau_d$. Since $\tau_d = \tau_{d,0} - t$ where $\tau_{d,0} = \tau_d(t=0)$, we have $u = -1$, and equation \ref{eq: continuity equation} reduces to 
\begin{equation}
\frac{\partial\rho}{\partial t} - \frac{\partial \rho}{\partial x} = S(x).
\label{eq: continuity equation for x = tau}
\end{equation}
Appendix \ref{sec: alternative derivation} presents an alternative derivation of equation \ref{eq: continuity equation for x = tau} that does not invoke a continuity equation and is a more direct parallel to the Section \ref{sec: toy model} experiments.

We will consider the general solution $\rho(x,t)$ in Section \ref{sec: PDE solution}, but let us first explore the case that the distribution is steady state ($\partial\rho/\partial t = 0$). The steady-state solution $\rho_{ss}(x)$ obeys
\begin{equation}
\frac{d\rho_{ss}}{dx} = -S(x) \ \ (\mathrm{steady \ state \ with \ } x = \tau_d).
\label{eq: drho/dx steady state for x = tau}
\end{equation}
This reduces to one dimension and can be solved directly for $\rho_{ss}(x)$ when the source distribution is known. In our first simulated population from the previous section, the planets were populated uniformly in semi-major axis and age. Since $\tau_d \propto a^{13/2}$ (equations \ref{eq: tau_a} and \ref{eq: tau_decay}), this implies that ${S(x) = A x^{-11/13}}$ with $A$ being a positive constant. Solving equation \ref{eq: drho/dx steady state for x = tau}, we find the steady-state solution 
\begin{equation}
\rho_{ss}(x) = C_1 - \frac{13}{2} A x^{2/13}, 
\label{eq: rho(x) steady state for x = tau}
\end{equation}
where $C_1$ is a positive constant. 



We now consider the case where $x = \ln\tau_d$. Here $u = -e^{-x}$, and equation \ref{eq: continuity equation} reduces to 
\begin{equation}
\frac{\partial\rho}{\partial t} - e^{-x}\left(\frac{\partial \rho}{\partial x} - \rho(x) \right) = S(x).
\label{eq: continuity equation for x = ln(tau)}
\end{equation}
The steady-state solution obeys 
\begin{equation}
\frac{d\rho_{ss}}{dx} - \rho_{ss}(x) = -S(x)e^x \ \ (\mathrm{steady \ state \ with \ } x = \ln\tau_d).
\label{eq: drho/dx steady state for x = ln(tau)}
\end{equation}
Again considering the case where the planets are populated uniformly in semi-major axis and age, the source distribution is $S(x) = A e^{2x/13}$, and the solution is 
\begin{equation}
\rho_{ss}(x) = C_1 e^x - \frac{13}{2} A e^{15x/13}.
\label{eq: rho(x) steady state for x = ln(tau)}
\end{equation}
This is equivalent to our earlier result (equation \ref{eq: rho(x) steady state for x = tau}) given that the transformation from $x = \tau_d$ to $x = \ln \tau_d$ requires introducing an additional factor of $e^x$.

We can proceed further with this steady-state solution. Taking the derivative of $\ln\rho(x)$ with respect to $x=\ln\tau_d$, we calculate the slope of the distribution in log-log space. This is equal to
\begin{equation}
\begin{split}
p \equiv \frac{d\ln\rho_{ss}(x)}{dx} &= \frac{1-(e^{x}/e^{x_p})^{2/13}}{1-\frac{13}{15}(e^{x}/e^{x_p})^{2/13}} \\
&= \frac{1-(\tau_d/\tau_{d,p})^{2/13}}{1-\frac{13}{15}(\tau_d/\tau_{d,p})^{2/13}}, 
\label{eq: dlnrho/dx}
\end{split}
\end{equation}
where $\tau_{d,p} = e^{x_p}$ is the location of the peak of the steady-state solution,
\begin{equation}
\tau_{d,p} = e^{x_p} = \left(\frac{2}{15}\frac{C_1}{A}\right)^{13/2}.
\label{eq: tau_dp}
\end{equation}
We call this slope $p$ because it is analogous to the exponent of the power law fitting function (equation \ref{eq: f(log10 tau)}) we used with the simulated distributions in Section \ref{sec: toy model}.
However, it is important to note that $d\ln\rho_{ss}(x)/dx$ is not constant with $\tau_d$, although it varies only marginally for $\tau_d<\tau_{d,p}$. This indicates that the power law fit in Section \ref{sec: toy model} is only an approximation, and the steady-state solution is not actually a power law, although it can be well-approximated by one for $\tau_d < \tau_{d,p}$. Equation \ref{eq: dlnrho/dx} evaluates to $p \rightarrow 1$ for $\tau_d/\tau_{d,p} \rightarrow 0$ (very short decay times) and $p = 0.84$ for $\tau_d/\tau_{d,p} = 0.03$. As expected, the latter value agrees very well with the slopes calculated for the simulated distributions (Figures \ref{fig: tau_decay distribution} and \ref{fig: tau_decay distribution with randomized params}), and 0.03 Gyr is the approximate midpoint of the steady-state region in those cases.

\subsection{General solution of $\rho(x,t)$}
\label{sec: PDE solution}

\begin{figure}
\centering
\includegraphics[width=\columnwidth]{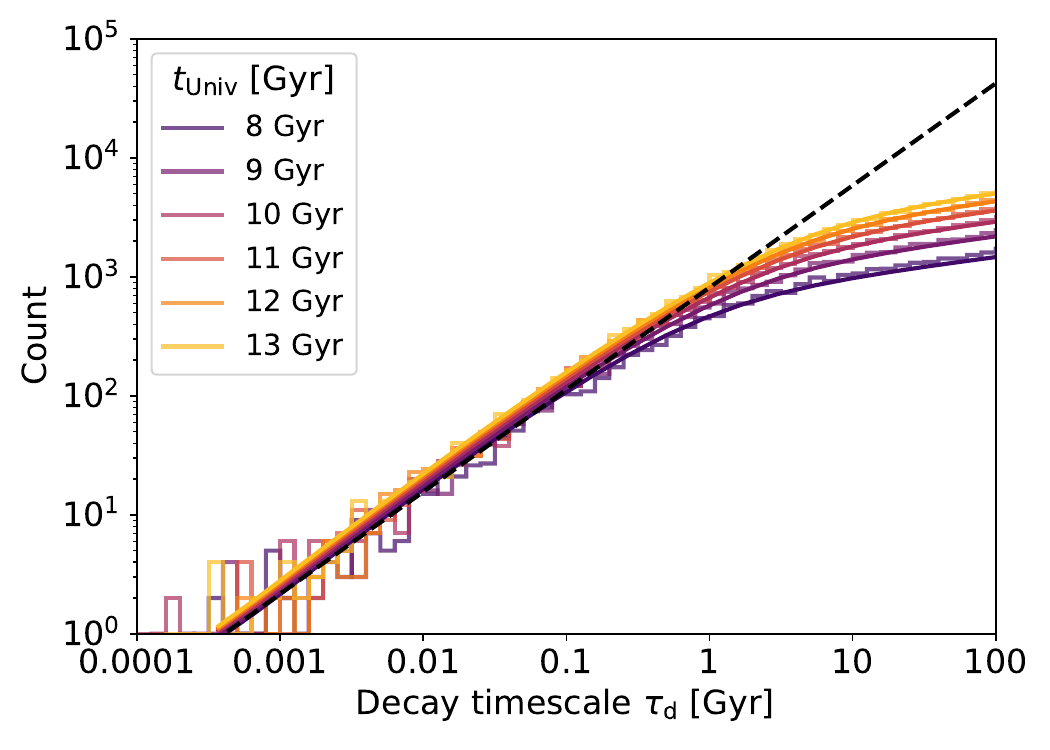}
\caption{Comparison of the numerical PDE solution of $\rho(x,t)$ and the simulated distributions of $\tau_d$ shown in Figure \ref{fig: tau_decay distribution}. The PDE solutions are the solid lines overlaid on the histograms. } 
\label{fig: simulated tau_decay distribution with PDE solution}
\end{figure}

Thus far we have only considered the steady-state solutions. Here we explore the general solution of $\rho(x, t)$ and show that the steady-state solution adequately describes the subset of the distribution with $\tau_d \lesssim t$. Let us consider equation \ref{eq: continuity equation for x = ln(tau)} for which $x = \ln\tau_d$. We set $S(x) \propto e^{2x/13}$ as before and $\rho(x,0) \propto S(x)$. We use the \texttt{py-pde}\footnote{https://py-pde.readthedocs.io/} partial differential equation (PDE) solver \citep{Zwicker2020} to numerically evaluate $\rho(x,t)$ at a set of defined time values. Here and later in the paper, we will call these curves of $\rho(x,t)$ the ``PDE solution''.

We first check the PDE solution relative to the simulated distributions from Section \ref{sec: toy model}. We use the first version of the simulated distributions (Figure \ref{fig: tau_decay distribution}). We obtain $\rho(x,t)$ at six times separated by 1 Gyr. The curves are then scaled (all by the same scale factor) such that they agree with the histograms from the simulated distributions. The resulting comparison between the simulated distributions and the numerical PDE solution is shown in Figure \ref{fig: simulated tau_decay distribution with PDE solution}. The agreement is excellent, and it is clear that the PDE solution is an accurate model for the simulation results. 

\begin{figure}
\centering
\includegraphics[width=\columnwidth]{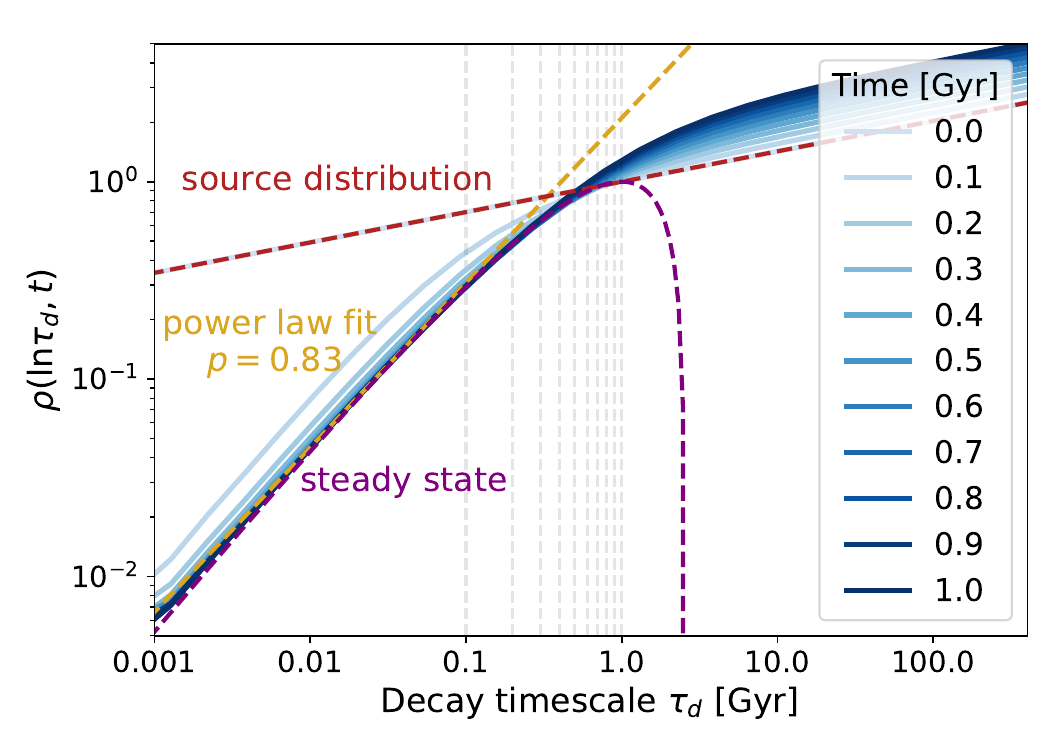}
\includegraphics[width=\columnwidth]{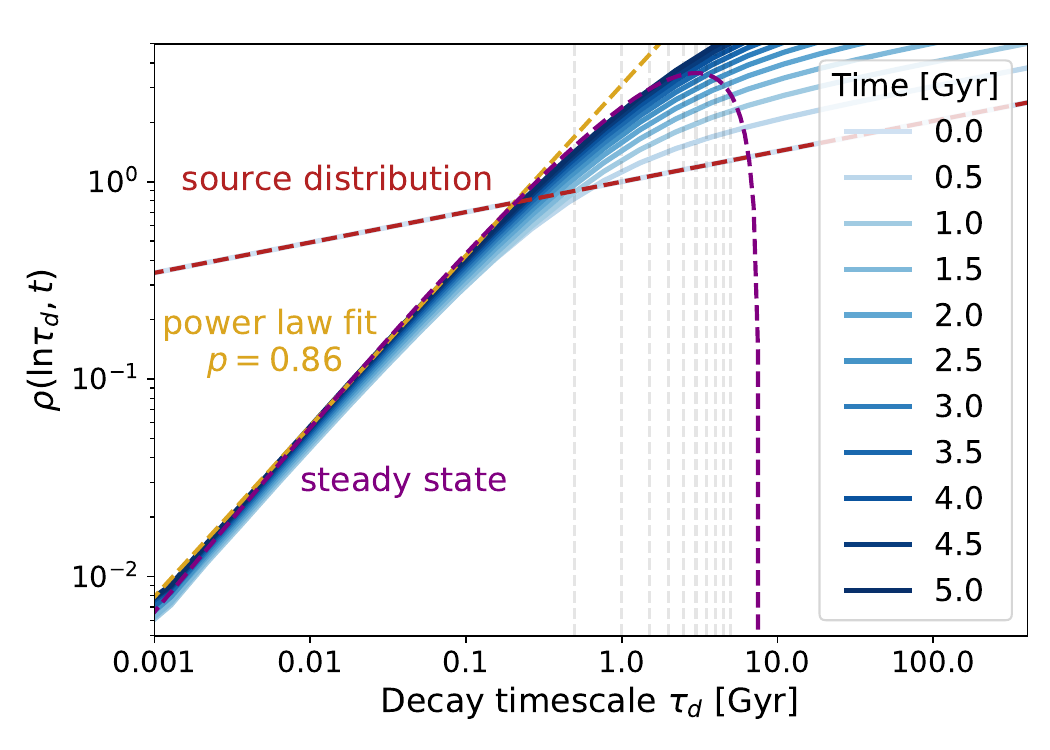}
\includegraphics[width=\columnwidth]{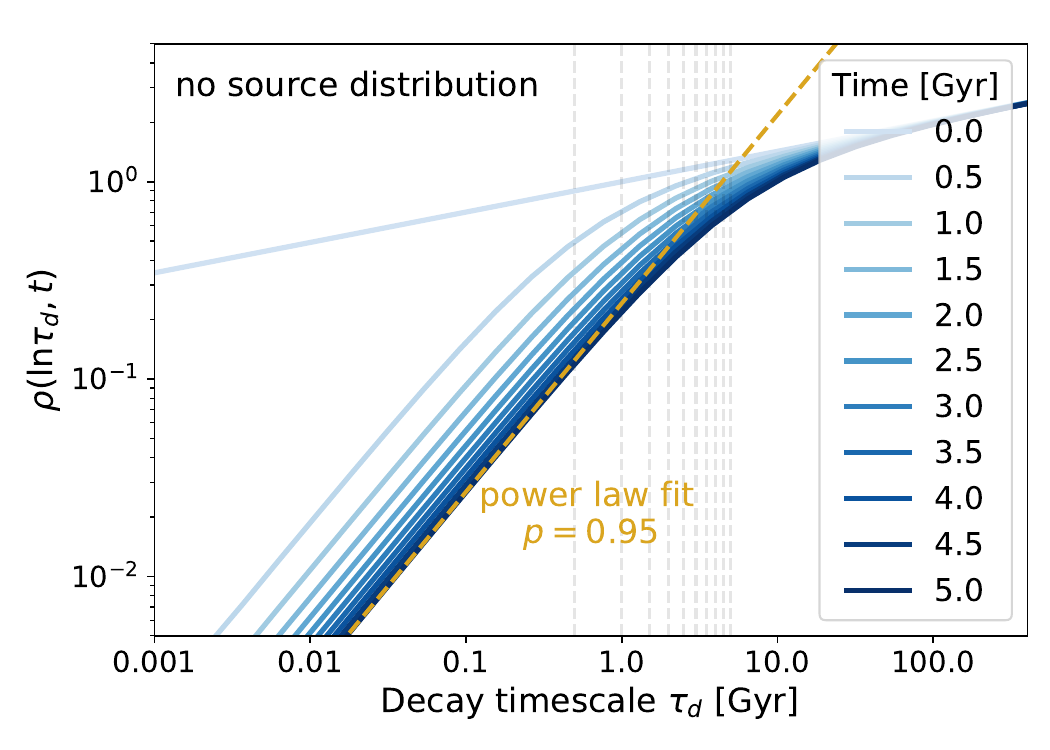}
\caption{Numerical PDE solution to equation \ref{eq: continuity equation for x = ln(tau)}. The top and middle panels use ${S(x) \propto e^{2x/13}}$ and ${\rho(x,t=0) \propto S(x)}$ where $x=\ln\tau_d$. They show the solution over a time period of 1 Gyr and 5 Gyr, respectively. The blue curves show the solution at each time indicated in the legend. The red dashed line shows the source distribution $S(x)$. The yellow dashed line shows a power law fit to the final blue curve within the range $\tau_d \sim 0.005-0.1$ Gyr. The purple dashed line shows the steady-state solution (equation \ref{eq: rho(x) steady state for x = ln(tau)}), which is clearly a good fit for $t \lesssim \tau_d$. The vertical dashed lines indicate each time. The bottom panel has no source distribution, ${S(x) = 0}$, but sets and ${\rho(x,t=0)}$ the same as the other panels.} 
\label{fig: PDE solution}
\end{figure}

It is worthwhile to explore the PDE solution in more detail and understand its general behavior. Figure  \ref{fig: PDE solution} shows the solution over time periods spanning either 1 Gyr or 5 Gyr. The solution is approximately a broken power law, with the location of the break being $\tau_d \sim t$. We plot the steady-state solution (equation \ref{eq: rho(x) steady state for x = ln(tau)}) with $C_1/A = 15/2\tau_{d,p}^{2/13}$ (equation \ref{eq: tau_dp}), which is a good approximation for $\tau_d<\tau_{d,p}\sim t$ but sharply deviates for $\tau_d>\tau_{d,p}$. It is intuitive that the steady-state solution is valid approximately up to the location of the break at $\tau_d \sim t$. Planets with $\tau_d \lesssim t$ have time to decay within the timescale of the evolution of the population, whereas the planet population with $\tau_d \gtrsim t$ just grows continuously since planets born with $\tau_d \gtrsim t$ take a long time to decay substantially. 

The final panel of Figure \ref{fig: PDE solution} shows the PDE solution with no source distribution, which corresponds to a case where all planets are born at one instant in time. The distribution shrinks over time but there is still a region that maintains an approximately constant slope with power law exponent $p = 0.95$. It is important to note that this slope is fairly similar to the earlier case with a non-zero source distribution. This further emphasizes the insensitivity of the steady-state region to the source distribution and implies that the slope is predominantly driven by the dynamics of tidal decay.

\section{Generalization to arbitrary tidal parameterizations}
\label{sec: generalization}

The constant $Q_{\star}'$ framework introduced in Section \ref{sec: toy model} results in a particular dependence of $\tau_d$ on $a$, specifically $\tau_d \propto a^{13/2}$. However, this simplest relation probably doesn't hold in reality, as most tidal dissipation mechanisms predict some dependence of $Q_{\star}'$ on the tidal frequency. Understanding this dependence allows us to constrain the dissipation mechanisms, as we will discuss later in Section \ref{sec: theoretical models}. In this section, we will introduce different ways to parameterize $\tau_d$. Before we get into specifics, let us first consider the possibility that $\tau_d \propto a^{1/\gamma}$ where $\gamma > 0$ is some positive constant. It has to be positive because otherwise $\tau_d \rightarrow \infty$ as $a \rightarrow 0$, which is nonphysical.

If $x=\ln\tau_d$ and the planets are still taken to be born uniformly in semi-major axis,\footnote{We use a uniform distribution here because it is simple and tractable for the analytic solution. Although more realistic initial semi-major axis distributions are not solvable analytically, Appendix \ref{sec: source distribution} presents numerical results for the simulated hot Jupiter populations with different choices for the initial semi-major axis distribution.} then $S(x)=A e^{\gamma x}$, where $A$ is a positive constant. The steady-state solution to equation \ref{eq: continuity equation for x = ln(tau)} is then 
\begin{equation}
\rho_{ss}(x) = C_1e^x - \frac{A}{\gamma}e^{(\gamma+1)x}.
\label{eq: rho(x) steady state for x = ln(tau) and arbitrary gamma}
\end{equation}
The slope in log-log space is 
\begin{equation}
\begin{split}
p = \frac{d\ln\rho_{ss}(x)}{dx} &= \frac{1-(e^x/e^{x_p})^{\gamma}}{1-\frac{1}{(\gamma+1)}(e^x/e^{x_p})^{\gamma}} \\
&= \frac{1-(\tau_d/\tau_{d,p})^{\gamma}}{1-\frac{1}{(\gamma+1)}(\tau_d/\tau_{d,p})^{\gamma}},
\label{eq: dlnrho/dx for arbitrary gamma}
\end{split}
\end{equation}
where $\tau_{d,p} = e^{x_p}$ is the location of the peak of the steady-state solution, 
\begin{equation}
\tau_{d,p} = e^{x_p} = \left(\frac{C_1}{A(1+1/\gamma)}\right)^{1/\gamma}.
\label{eq: tau_dp for x = ln(tau) and arbitrary gamma}
\end{equation}
Equations \ref{eq: rho(x) steady state for x = ln(tau) and arbitrary gamma} through \ref{eq: tau_dp for x = ln(tau) and arbitrary gamma} agree with our earlier derivation (equations \ref{eq: rho(x) steady state for x = ln(tau)} through \ref{eq: tau_dp}) when $\gamma = 2/13$.


\subsection{\cite{2018AJ....155..165P} parameterization}
\label{sec: P18 parameterization}

\cite{2018AJ....155..165P} derived empirical constraints on $Q_{\star}'$ by considering the tidal spin-up of hot Jupiter host stars. The rotational spin-up is the complement to the tidal orbital decay we have been exploring in this work; that is, the angular momentum lost from the planet's orbit is transferred into the rotation of the host star. Considering a sample of 188 hot Jupiter systems, \cite{2018AJ....155..165P} calculated the values of $Q_{\star}'$ that generate enough spin-up to explain the stars' present-day observed rotation rates, starting from initial conditions dictated by observations of single stars within young clusters. They found a sharp dependence of the resulting $Q_{\star}'$ constraints on the tidal forcing frequency. Their results were well described by the fitting function
\begin{equation}
Q_{\star}'(P_{\mathrm{tide}}) = \max\left[10^6\left(\frac{P_{\mathrm{tide}}}{\mathrm{days}}\right)^{-3.1}, \ 10^5\right],
\label{eq: Penev Q_star' fitting function}
\end{equation}
where $P_{\mathrm{tide}}$ is the tidal forcing period. This is half of the orbital period of the planet in a reference frame rotating with the stellar spin,
\begin{equation}
P_{\mathrm{tide}} \equiv \frac{1}{2(P^{-1} - P_{\mathrm{rot}}^{-1})} \approx \frac{P}{2},
\end{equation}
where $P$ is the orbital period of the planet and $P_{\mathrm{rot}}$ is the rotational period of the star, and the approximation holds when $P \ll P_{\mathrm{rot}}$.

Motivated by \cite{2018AJ....155..165P}'s findings, we here consider a parameterization 
\begin{equation}
Q_{\star}'(P) = \max\left[Q_0\left(\frac{P} {2 \ \mathrm{days}}\right)^{\alpha}, 10^4\right],
\label{eq: Penev Q_star' with P approximation}
\end{equation}
where we will aim to derive constraints on $\alpha$ and $Q_0$. We use a lower minimum value of $Q_{\star}'$ to allow greater flexibility of the model, and we assume $P \ll P_{\mathrm{rot}}$. We note that the parameterization in equation \ref{eq: Penev Q_star' with P approximation} is very simple and ignores other dependencies that probably matter at some level, such as variations with stellar mass and time. We believe this simple approach is warranted as a first attempt towards deriving constraints from the decay timescale distribution, but more complicated models should be explored in the future. See Section \ref{sec: caveats and extensions} for further discussion.

For the part of the $Q_{\star}'$ function in equation \ref{eq: Penev Q_star' with P approximation} that goes as a power law, the instantaneous orbital decay rate is then modified from equation \ref{eq: tau_a unit version} to  
\begin{equation}
\frac{1}{\tau_a} = \frac{C}{Q_0}\left(\frac{M_{\star}}{M_{\odot}}\right)^{\frac{\alpha-1}{2}}\left(\frac{R_{\star}}{R_{\odot}}\right)^5\left(\frac{M_p}{M_{\mathrm{Jup}}}\right)\left(\frac{a}{\mathrm{AU}}\right)^{-\frac{13+3\alpha}{2}}
\label{eq: tau_a Penev version}
\end{equation}
where 
\begin{equation}
C = -\frac{9}{2}\frac{(\mathrm{days})^{\alpha}}{\pi^{\alpha}}\frac{G^{\frac{\alpha+1}{2}}R_{\odot}^5 M_{\mathrm{Jup}}}{{M_{\odot}}^{\frac{1-\alpha}{2}}(\mathrm{AU})^{\frac{13+3\alpha}{2}}}.
\label{eq: C Penev version}
\end{equation}
This parameterization yields ${\tau_d \propto a^{(13+3\alpha)/2}}$, so $\gamma = 2/(13+3\alpha)$. Since $\gamma > 0$, this implies that $\alpha$ is limited to the range $\alpha > -4.33$. The tidal decay timescale $\tau_d$ is related to $\tau_a$ by 
\begin{equation}
\tau_d = - \gamma\tau_a = -\frac{2}{13+3\alpha}\tau_a,
\label{eq: tau_decay with variable alpha}
\end{equation}
which reduces to equation \ref{eq: tau_decay} when $\alpha = 0$.


\subsection{Dynamical timescale parameterization}
\label{sec: dynamical timescale parameterization}

The \cite{2018AJ....155..165P} parameterization does not incorporate stellar properties into the function for $Q_{\star}'$. An alternative parameterization is to include the dynamical timescale of the star,
\begin{equation}
t_{\star} = \sqrt{\frac{R_{\star}^3}{G M_{\star}}},
\label{eq: t_star}
\end{equation}
which is $\sim 0.44$ hr for the Sun. In this case, we are scaling the characteristic timescale to that of the dissipating body, the star. In the case of dynamical tidal oscillations, damping might be proportional to typical oscillatory periods, which will scale with the stellar dynamical time. Again working with the approximation that $P \ll P_{\mathrm{rot}}$, we can consider the parameterization
\begin{equation}
Q_{\star}'(P) = \max\left[Q_0\left(\frac{P}{100 \ t_{\star}}\right)^{\alpha}, 10^4\right].
\label{eq: Q_star' with stellar dynamical timescale}
\end{equation}
The factor of 100 is necessary such that the fraction within parentheses is of order unity. A comparison of this parameterization to that of equation \ref{eq: Penev Q_star' with P approximation} allow us to test whether there is evidence for tidal dissipation scaling with stellar properties in addition to orbital period. For the part of the $Q_{\star}'$ function that goes as a power law, the instantaneous orbital decay rate is then equal to
\begin{equation}
\frac{1}{\tau_a} = \frac{C}{Q_0}\left(\frac{M_{\star}}{M_{\odot}}\right)^{-\frac{1}{2}}\left(\frac{R_{\star}}{R_{\odot}}\right)^{\frac{3\alpha+10}{2}}\left(\frac{M_p}{M_{\mathrm{Jup}}}\right)\left(\frac{a}{\mathrm{AU}}\right)^{-\frac{13+3\alpha}{2}}
\label{eq: tau_a t_star version}
\end{equation}
where
\begin{equation}
C = -\frac{9}{2}\frac{100^{\alpha}}{(2\pi)^{\alpha}}\frac{G^{\frac{1}{2}}R_{\odot}^{\frac{3\alpha+10}{2}}M_{\mathrm{Jup}}}{M_{\odot}^{\frac{1}{2}}(\mathrm{AU})^{\frac{13+3\alpha}{2}}}.
\end{equation}
We note that this parameterization still results in the same semi-major axis dependence as the \citetalias{2018AJ....155..165P} parameterization, ${\tau_d \propto a^{(13+3\alpha)/2}}$, and the same relation in equation \ref{eq: tau_decay with variable alpha}.

\section{Observations}
\label{sec: observations}

\subsection{Planet sample} 

We now compare the expectations from the tidal models to the observed planet population in an effort to constrain its properties. Specifically, we aim to constrain $\alpha$ and $Q_0$. We construct a sample of observed planets from \cite{PSCompPars} \citep{2013PASP..125..989A}. We use the composite parameters table of planetary systems, considering only planets for which the mass was measured. We also limit to planets with small eccentricities, $e < 0.02$, so as to restrict ourselves to systems for which stellar tides dominate over planetary tides. We limit the sample to transiting planets for reasons that we will soon clarify. Out of an initial 5,470, this yields a sample of 655 planets, which we call the ``all planets sample'' or ``AP sample''. We also create a sample of hot Jupiters by considering those with measured masses between $0.5 \ M_{\mathrm{Jup}}$  and $13 \ M_{\mathrm{Jup}}$ and periods between 0.1 days and 10 days. This yields a sample of 252 hot Jupiters, which we call the ``HJ sample''. 

With the measured properties of the observed systems, we have everything we need to evaluate $\tau_d$ except for $Q_0$ and $\alpha$. We will thus vary $Q_0$ and $\alpha$ and compare the $\tau_d$ distribution of the observed planet sample with the simulated analogs discussed earlier. Given that the $\tau_d$ distribution depends on various planet properties, it is vital to account for selection effects in the observed population. This is why we restrict the sample to transiting planets only. Then, whenever we compute $\tau_d$ histograms in the upcoming analysis, we weight them by the inverse of the transit probability, $P_{\mathrm{trans}} \approx (R_{\star} + R_p)/a$.


\subsection{Constant $Q_{\star}'$}
\label{sec: uniform Qstar}

\begin{figure}[t!]
\centering
\includegraphics[width=\columnwidth]{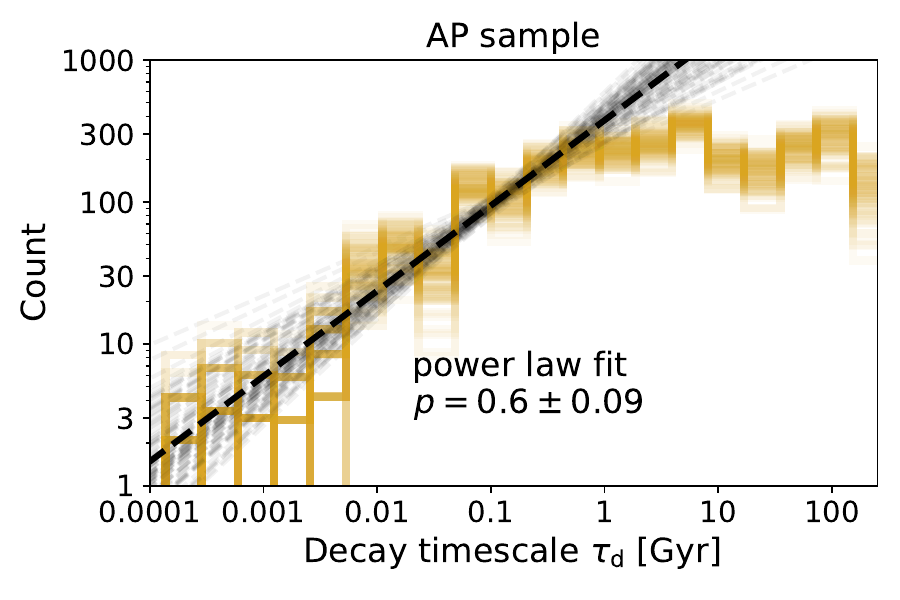}
\includegraphics[width=\columnwidth]{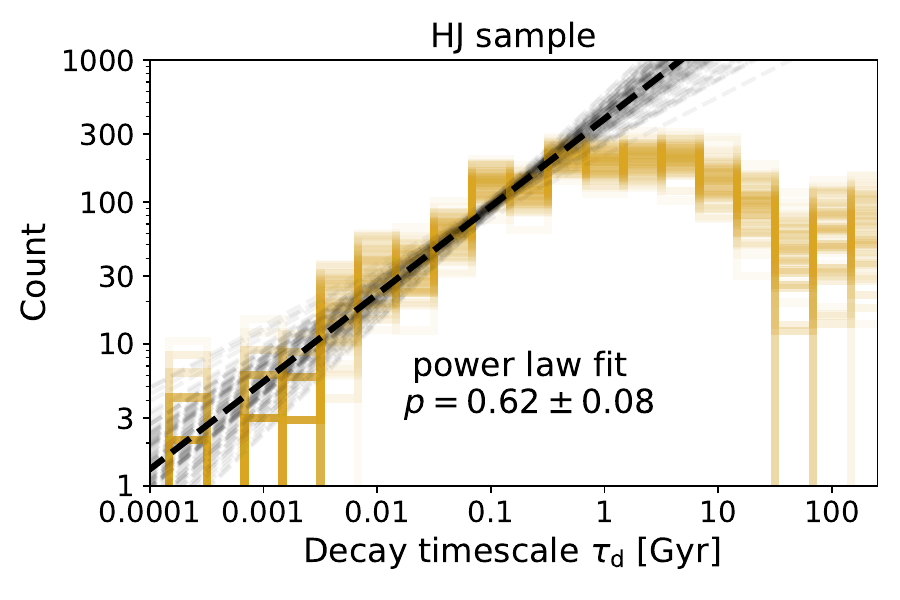}
\caption{Distributions of $\tau_d$ for the all planets (AP) sample and hot Jupiters (HJ) sample assuming constant $Q_{\star}'$ (such that ${\alpha = 0}$, $\gamma = 2/13$). A bootstrap resampling method is used to generate 100 histograms with 100 corresponding power law fits to the region $0.001 \ \mathrm{Gyr} < \tau_d < 0.3 \ \mathrm{Gyr}$. The dark black dashed line indicates the mean power law fit. The mean and standard deviation of the power law exponent are reported.} 
\label{fig: observed distribution (all planets and hot jupiters) with alpha = 0}
\end{figure}

\begin{figure*}[t!]
\centering
\includegraphics[width=\textwidth]{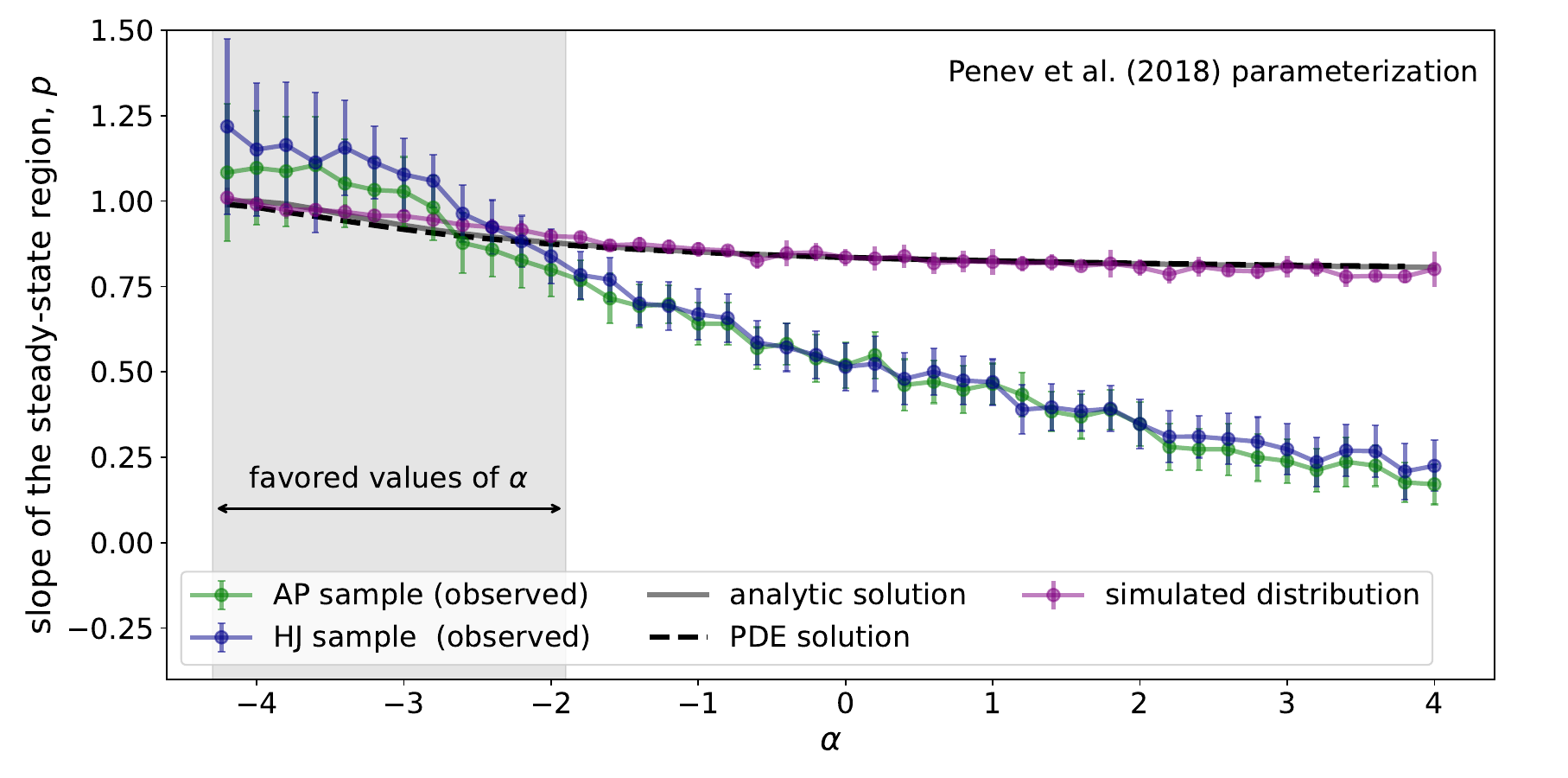}
\includegraphics[width=\textwidth]{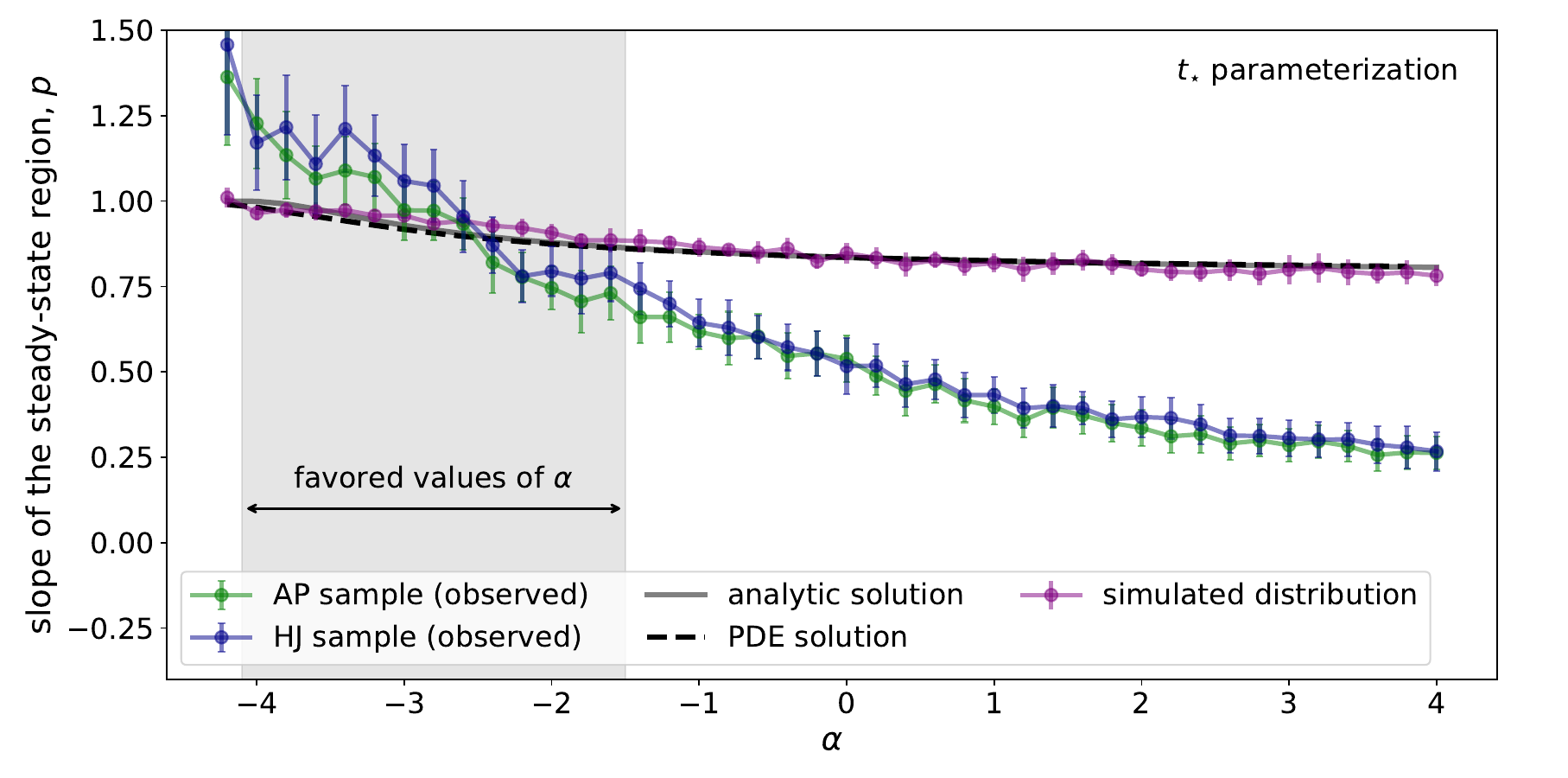}
\caption{Slope of the steady-state region (equal to the exponent of the power law fit) as a function of $\alpha$. The top panel corresponds to the \citetalias{2018AJ....155..165P} parameterization of $Q_{\star}'$ (equation \ref{eq: Penev Q_star' with P approximation}), while the bottom panel corresponds to the stellar dynamical parameterization (equation \ref{eq: Q_star' with stellar dynamical timescale}).  With $\gamma = 2/(13+3\alpha)$, the gray solid line shows the analytic slope (equation \ref{eq: dlnrho/dx for arbitrary gamma}) with $\tau_{d,p} = 1$ Gyr and $\tau_d = 0.03$ Gyr (which is approximately halfway between 0.001 Gyr and 1 Gyr in log space). The dashed black line shows a numerical derivative ($d\ln\rho(x)/dx$) of the PDE solution to equation \ref{eq: continuity equation for x = ln(tau)} evaluated at $\tau_d = 0.03$ Gyr. The purple curve indicates the results for the simulated distributions, where we show the slopes of a power law fit to the steady-state region with $\tau_d < 1$ Gyr. The points and errorbars represent the mean and standard deviation from 10 simulated distributions. The analytic solution, PDE solution, and simulated distributions agree for all $\alpha$, as expected. The green and blue curves indicate the results for the observed all planets (AP) sample and hot Jupiters (HJ) sample. The points and errorbars represent the mean and standard deviation from 100 bootstrap trials. The shaded gray region indicates the favored values of $\alpha$ for the observed planet population based on the agreement of the slope between the observed and simulated, analytic, and numerical solutions.\\ \\} 
\label{fig: slope_vs_alpha}
\end{figure*}

\begin{figure}[t!]
\centering
\includegraphics[width=\columnwidth]{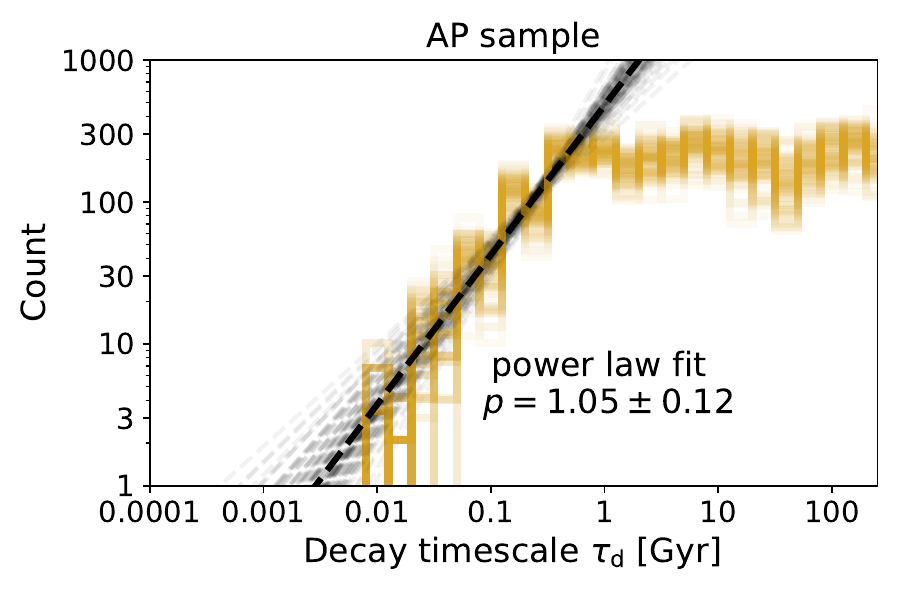}
\includegraphics[width=\columnwidth]{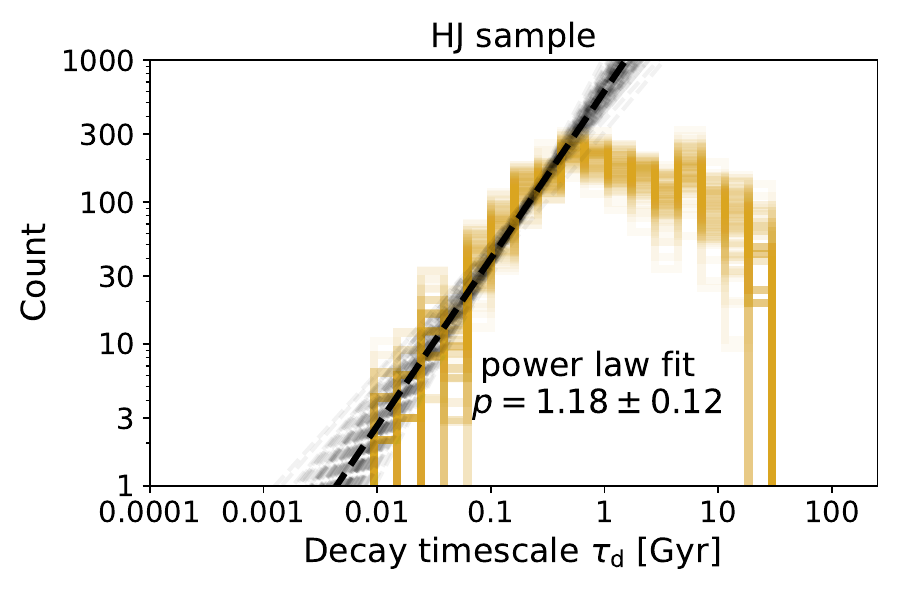}
\caption{Distributions of $\tau_d$ for the all planets (AP) sample and hot Jupiters (HJ) sample using the \citetalias{2018AJ....155..165P} parameterization and assuming $\alpha=-3$ and $Q_0 = 10^6$. A bootstrap resampling method is used to generate 100 histograms with 100 corresponding power law fits to the region $0.001 \ \mathrm{Gyr} < \tau_d < 0.4 \ \mathrm{Gyr}$. The dark black dashed line indicates the mean power law fit. The mean and standard deviation of the power law exponent are reported.} 
\label{fig: observed distribution (all planets and hot jupiters) with alpha = -3}
\end{figure}

We first plot the distributions of decay times for the AP sample and HJ sample assuming the simple case of no tidal frequency dependence on $Q_{\star}'$ (so $\alpha = 0$) and a constant $Q_{\star}' = Q_0 = 10^6$ for all stars. To capture the spread in the distribution, we apply a bootstrapping method by sampling with replacement from the observed $\tau_d$ distribution to create 100 new $\tau_d$ distributions, each with the same size as the original one. We then create 100 histograms and perform a power law fit (equation \ref{eq: f(log10 tau)}) to the region $0.001 \ \mathrm{Gyr} < \tau_d < 0.3 \ \mathrm{Gyr}$ for each one. 

Figure \ref{fig: observed distribution (all planets and hot jupiters) with alpha = 0} shows the resulting distributions for the population of all planets and hot Jupiters. Interestingly, the region with $\tau_d \lesssim 1$ Gyr appears to be fairly well-described by a power law, as expected from our earlier simulations and numerical solutions. The fact that the observations show this power law dependence is an encouraging sign that we are on the right track. However, the slope is too shallow, falling around $p\sim0.6$ rather than $p\sim0.9$ seen earlier. This yields our first important conclusion; the observed distributions do not appear to be well-described by the default constant $Q_{\star}'$ model and must require some extra dependence of $Q_{\star}'$ on the planetary and/or stellar properties. 

\subsection{Constraints on $\alpha$ with \citetalias{2018AJ....155..165P} parameterization}
\label{sec: results using P18 parameterization}

Instead of a constant $Q_{\star}'$ model, we can see whether one of the alternative tidal parameterizations is a better fit. In this section we consider the \citetalias{2018AJ....155..165P} parameterization (equation \ref{eq: Penev Q_star' with P approximation}). We first explore the constraints on $\alpha$. The key idea is to determine how the slope of the steady-state region varies as a function of $\alpha$. If the observational data is well-described by a particular value of $\alpha$, then the slope of the steady-state region for the observed distributions should agree with theoretical distributions (both the simulated distributions and numerical PDE solution). Until Section \ref{sec: empirical constraints on Qstar'}, we keep a fixed $Q_0 = 10^6$. This will not affect our inferences because only $\alpha$ changes the \textit{slope} of the steady-state region, whereas $Q_0$ systematically shifts the whole $\tau_d$ distribution to larger or higher values (more on this in Section \ref{sec: empirical constraints on Qstar'}).

The results of these calculations are shown in the top panel of Figure \ref{fig: slope_vs_alpha}. For each $\alpha$ value, we compute an analytic, numerical, simulated, and observed measure of the slope. Specifically, we use: (1) the analytic value of the slope from equation \ref{eq: dlnrho/dx for arbitrary gamma} using $\tau_{d,p} = 1$ Gyr and $\tau_d = 0.03$ Gyr (where 0.03 Gyr is taken because it is approximately the log-space midpoint of the steady-state region), (2) a numerical calculation of the slope from the PDE solution to equation \ref{eq: continuity equation for x = ln(tau)} evaluated at $\tau_d = 0.03$ Gyr, (3) the slope values from the simulated distributions (based on power law fits), and (4) the slope values from the observed distributions. For the simulated distributions, we use the case with a uniform sampling of the semi-major axis,\footnote{Appendix \ref{sec: source distribution} considers other sampling choices for the semi-major axis.} $a/{\mathrm{AU}}\sim\mathcal{U}[0.01,0.1]$, and we use the mean and standard deviation of the power law fits.  For the slope values from the observed distributions, we calculate the estimates and errorbars based on a bootstrap resampling method. For each value of $\alpha$, we generate 100 histograms with 100 corresponding power law fits and then calculate the mean and standard deviation. 

We first note that all of the curves show that the slope decreases with increasing $\alpha$. This can be understood as follows. When $\alpha$ increases, $\gamma = 2/(13+3\alpha)$ decreases, and $\tau_d \propto a^{1/\gamma}$ becomes a steeper function of $a$. The $\tau_d$ distribution then spans a broader range of values and becomes shallower.  A second thing to note is that the curves of $p$ vs. $\alpha$ for the observed distributions show similar dependence between the two observational samples (all planets and hot Jupiters), but these curves differ from those of the analytic, numerical, and simulated distribution calculations. This is expected because not every value of $\alpha$ should be a good description of the observations. In other words, if there was agreement between the observational and theoretical curves across all values of $\alpha$, then there would be no possible constraint on $\alpha$.

The $p$ curves for the observed distributions only agree within uncertainties with the expected values for $\alpha$ values in the interval $\sim[-4.33, -2]$. This interval is noteworthy because it is consistent with \citetalias{2018AJ....155..165P}'s empirical result of $\alpha = -3.1$ as a best-fit value. This agreement is particularly meaningful considering that we are using completely independent methods for constraining $\alpha$. \citetalias{2018AJ....155..165P} considered the tidal rotational spin-up of the hot Jupiter host stars, whereas we are considering the distribution of tidal decay times of the planets. Both of these results suggest that the stellar tidal quality factors for interactions between stars and hot Jupiters go approximately as $Q_{\star}' \propto P^{-3}$.

The agreement of our $\alpha$ constraints with \citetalias{2018AJ....155..165P} motivates us to examine the full $\tau_d$ distribution for the case of $\alpha = -3$. Figure \ref{fig: observed distribution (all planets and hot jupiters) with alpha = -3} shows the $\tau_d$ distributions for the AP sample and HJ sample assuming $\alpha = -3$. It is clear to see that the distributions for each planet sample are cleanly described by a power law at the low $\tau_d$ end, and the slope is steeper than the uniform $Q_{\star}'$ case we examined in Section \ref{sec: uniform Qstar}. The slope is $p \sim 1.0$, and it agrees within uncertainties with the theoretical expectation of $p$ (as shown in Figure \ref{fig: slope_vs_alpha}). The distributions also show a much cleaner break location than the uniform $Q_{\star}'$ case examined earlier. 

\subsection{Constraints on $\alpha$ with the dynamical timescale parameterization}
\label{sec: results using dynamical timescale parameterization}

We repeat the same procedure as discussed in Section \ref{sec: results using P18 parameterization} to create the $p$ vs. $\alpha$ curves for the $t_{\star}$ parameterization. We show the results in the bottom panel of Figure \ref{fig: slope_vs_alpha}. Overall, the curves of $p$ vs. $\alpha$ for the observed distributions are similar to the the earlier case using the \citetalias{2018AJ....155..165P} parameterization. The favored $\alpha$ range is consistent with the earlier results. Thus, we do not see a strong preference in the current dataset for this model as opposed to the \citetalias{2018AJ....155..165P} parameterization. 

\subsection{Differences for hot stars and cool stars}
\label{sec: hot/cool stars}

\begin{figure*}[t!]
\centering
\includegraphics[width=\textwidth]{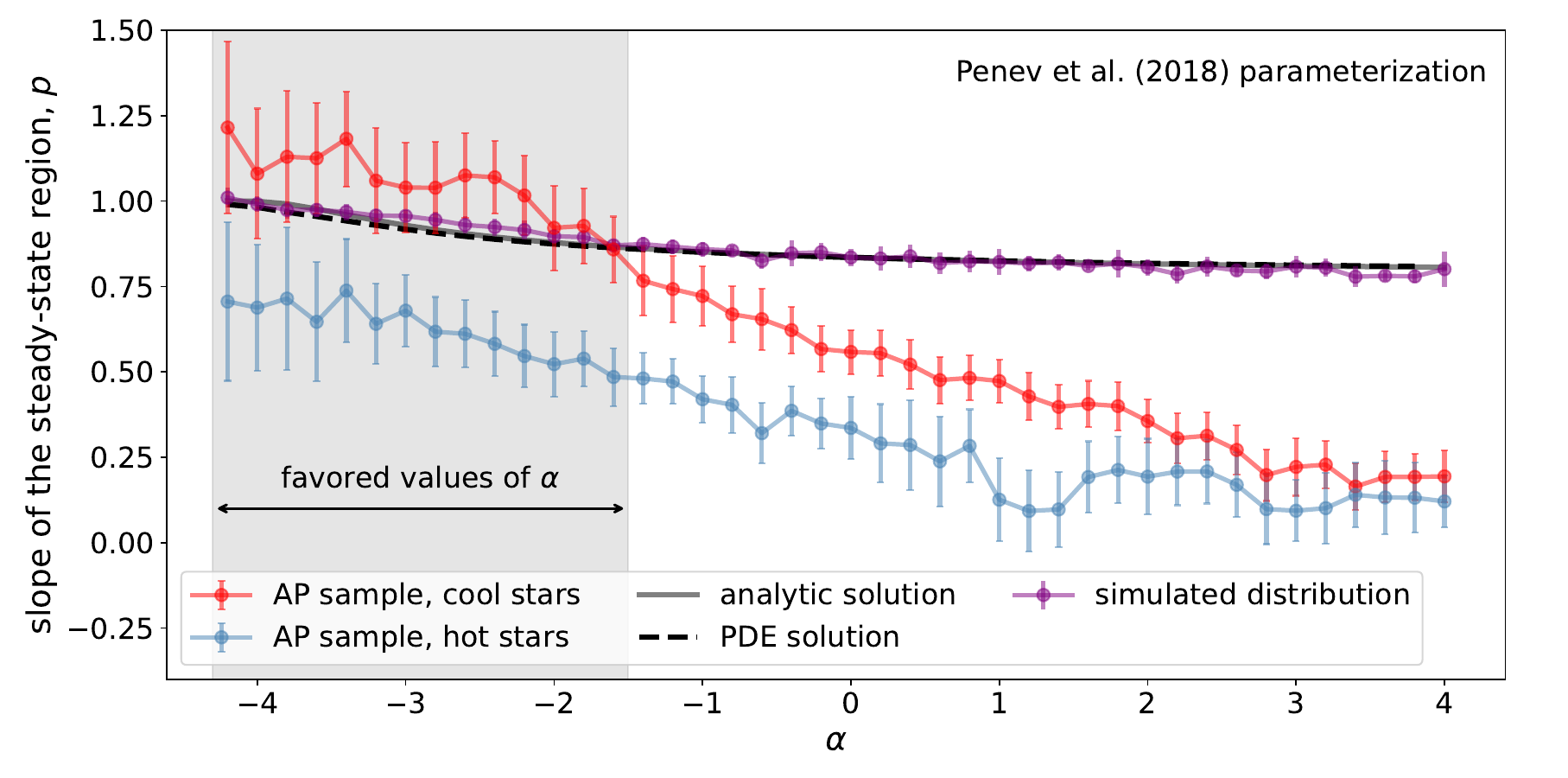}
\caption{Slope of the steady-state region as a function of $\alpha$ and split into the cool star and hot star samples. We use the \citetalias{2018AJ....155..165P} parameterization of $Q_{\star}'$. The curves for the analytic solution, PDE solution, and simulated distribution are the same as in Figure \ref{fig: slope_vs_alpha} (see caption for description). The red and blue curves indicate the results for the AP sample, with red/blue corresponding to the planets around cool/hot stars. The points and errorbars represent the mean and standard deviation from 100 bootstrap trials. The shaded gray region indicates the favored values of $\alpha$ for the observed planet population based on the agreement of the slope between the observed and simulated, analytic, and numerical solutions.} 
\label{fig: slope_vs_alpha_all_planets_cool_vs_hot}
\end{figure*}

\begin{figure}
\centering
\includegraphics[width=\columnwidth]{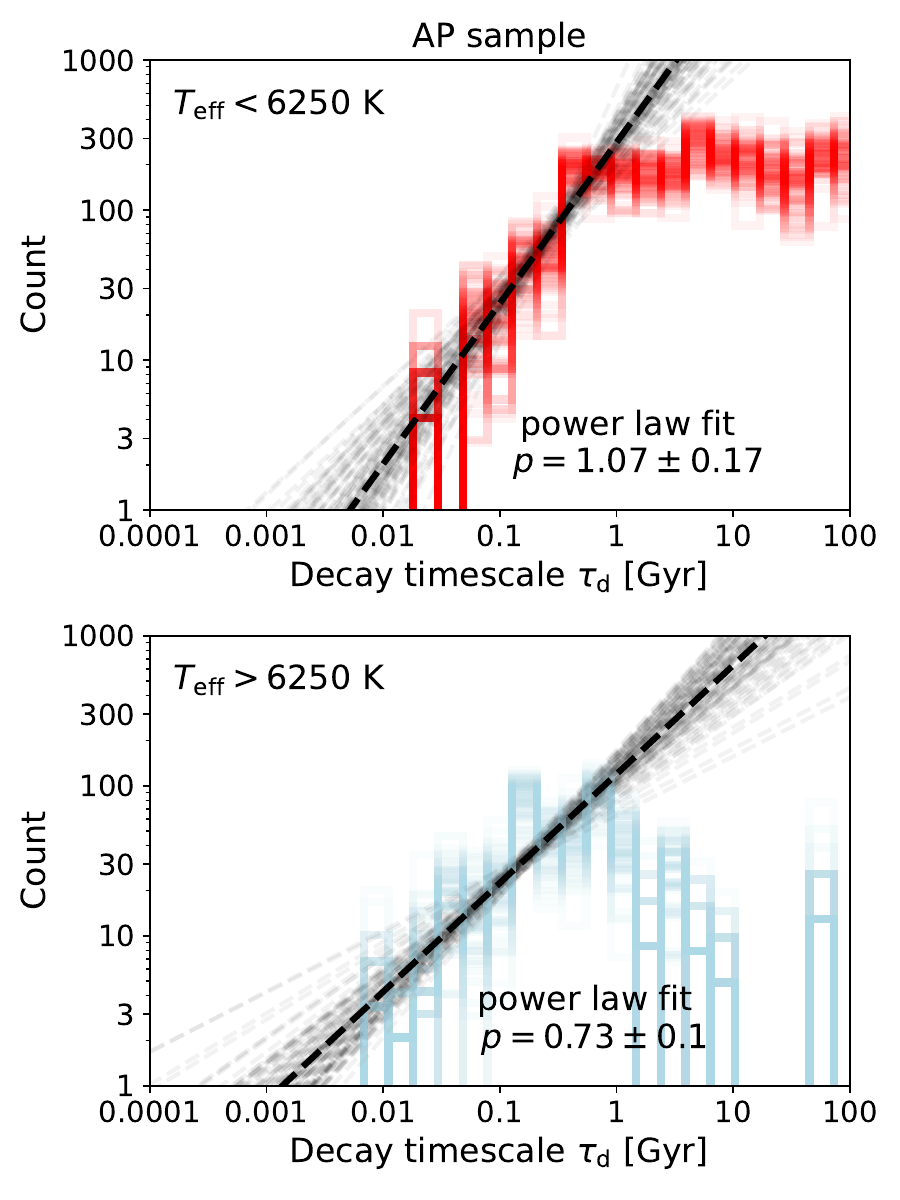}
\caption{Distributions of $\tau_d$ for the all planets (AP) sample assuming $\alpha=-3$ and $Q_0 = 10^6$. The top and bottom panels show systems below and above the Kraft break at $T_{\mathrm{eff}} = 6250$ K. A bootstrap resampling method is used to generate 100 histograms with 100 corresponding power law fits to the region $0.001 \ \mathrm{Gyr} < \tau_d < 0.4 \ \mathrm{Gyr}$. The dark black dashed line indicates the mean power law fit. The mean and standard deviation of the power law exponent are reported.} 
\label{fig: observed distribution: all planets}
\end{figure}

\begin{figure}
\centering
\includegraphics[width=\columnwidth]{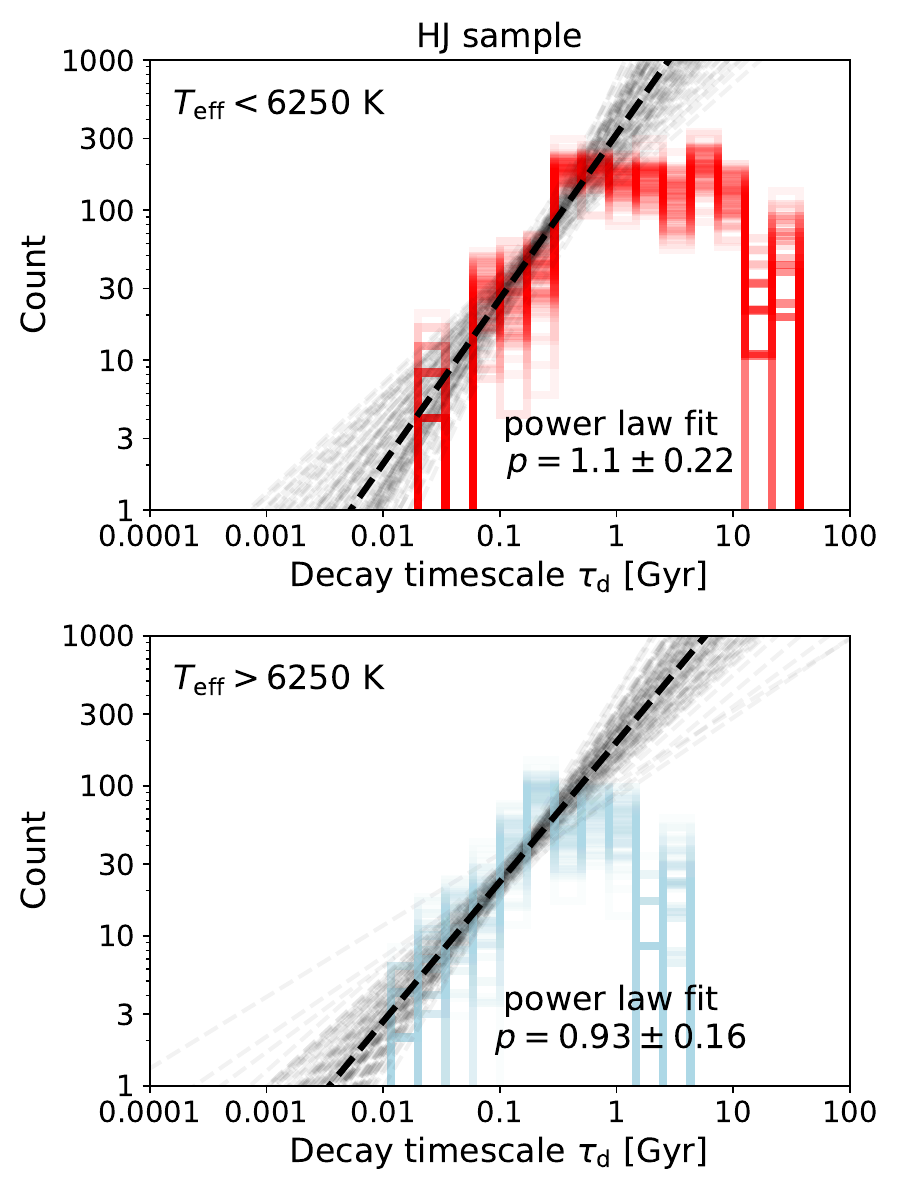}
\caption{Same as Figure \ref{fig: observed distribution: all planets} except for the HJ sample.} 
\label{fig: observed distribution: hot Jupiters}
\end{figure}

So far we have grouped our planet sample only by planet type. However, stellar type also plays a role in the strength of tidal interactions. For instance, it was established over a decade ago that hot Jupiters orbiting cool stars have stellar spin vectors that are almost always aligned with the planetary orbit normal vectors, whereas hot Jupiters orbiting hot stars have a broad range of stellar spin-orbit misalignments \citep{2010ApJ...718L.145W}. This is thought to be evidence that cool stars become realigned through tidal dissipation in their interiors \citep[e.g.][]{2012MNRAS.423..486L, 2014ApJ...784...66X, 2014ApJ...790L..31D, 2022ApJ...927...22S}. The transition temperature is approximately $T_{\mathrm{eff}} = 6250$ K, which corresponds to the ``Kraft break'' (named after \citealt{1967ApJ...150..551K}), below which cool stars have deep convective envelopes on the main sequence and undergo magnetic braking. The different internal structures of the stars with $T_{\mathrm{eff}} < 6250$ K could trigger stronger tidal dissipation, not only for stellar obliquity realignment but also for planetary orbital decay. 

Here we probe for differences in the $\tau_d$ distribution for planets orbiting hot stars and cool stars. Figure \ref{fig: slope_vs_alpha_all_planets_cool_vs_hot} shows the $p$ vs. $\alpha$ curves for the AP sample divided up into planets around cool stars and hot stars. Here we are using the \citetalias{2018AJ....155..165P} parameterization, but the $t_{\star}$ parameterization produces similar results and is thus not shown. Distinct differences are seen between the $p$ vs. $\alpha$ curves for the cool stars and the hot stars, and the differences are exactly in line with our expectations. First, the two curves barely overlap, and the hot stars curve does not overlap with theoretical expectations for any value of $\alpha$. This suggests that the hot star sample has not had sufficient tidal sculpting for there to be a steady-state portion of the $\tau_d$ distribution. We discuss this interpretation further below. For the cool star sample, on the other hand, the curve cleanly overlaps with theoretical expectation in a range of $\alpha$ that is wider but consistent with earlier results, $\alpha \in [-4.33, -1.5]$. 

The differences between the hot star and cool star samples are easier to visualize in the $\tau_d$ distribution. Figures \ref{fig: observed distribution: all planets} and \ref{fig: observed distribution: hot Jupiters} show the $\tau_d$ distributions for the AP sample and HJ sample assuming $\alpha = -3$. The different panels show sub-samples based on $T_{\mathrm{eff}}$. The $\tau_d$ distribution for the cool star sample shows a clean broken power law, just like theoretical expectations. The region with small decay times is well-fit by a steep power law with $p \sim 1.0$. This agrees with the theoretical expectations of the steady-state region when $\alpha \sim -3$. 

In contrast, the $\tau_d$ distribution for the hot star sample does not demonstrate the same steady-state features. Specifically, the region with small decay times does not appear to be well-described by a power law. We quantify this by computing the Pearson correlation coefficient, $r$, of the power-law fits 
as a measure of the goodness of fit. Averaging across bootstrapped trials and $\alpha \in [-4.33, -1.5]$, we find that $r = 0.93 \pm0.01$ for the cool star sample, while $r = 0.82 \pm 0.04$ for the hot star sample. The smaller $r$ value for the hot star sample indicates a poorer fit. Even if we could convince ourselves that a power law could describe the data, a secondary piece of evidence against it is that the slope is much shallower, $p \sim 0.7$. This disagrees with expectations of the steady-state region when $\alpha \sim -3$, as we saw in Figure \ref{fig: slope_vs_alpha_all_planets_cool_vs_hot}.

\subsection{Constraints on $Q_0$}

\begin{figure}[t!]
\centering
\includegraphics[width=\columnwidth]{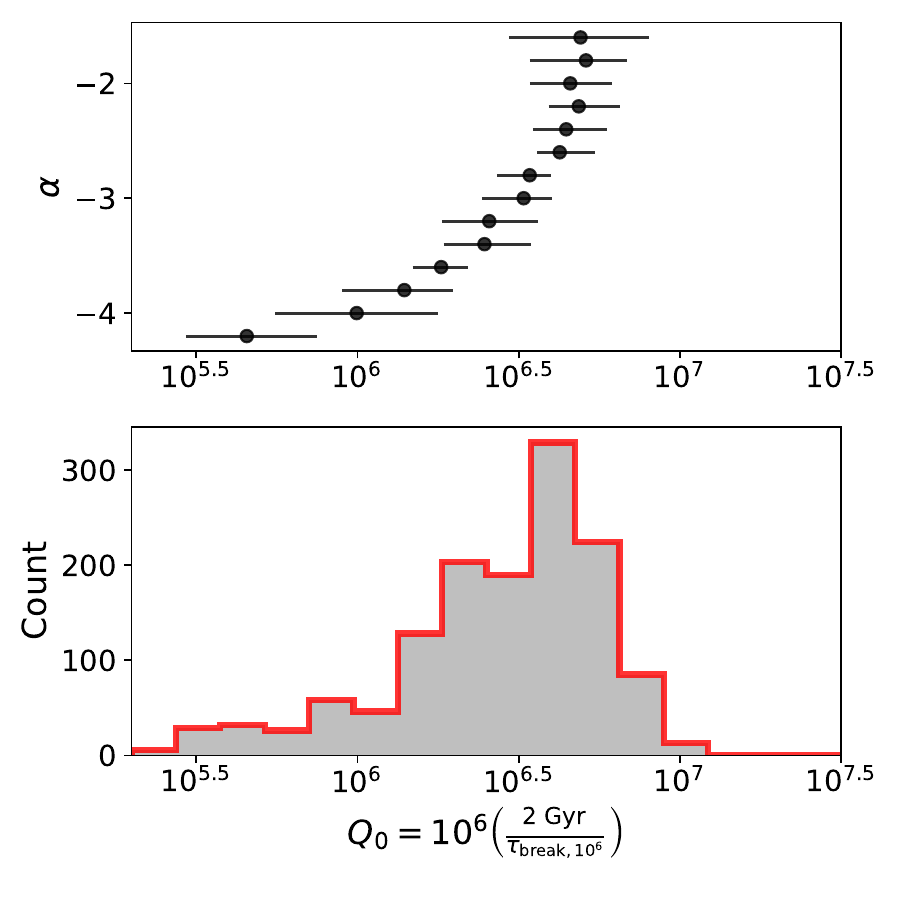}
\caption{Constraints on $Q_0$ from the AP sample around cool stars. The top panel shows the mean $Q_0$ for each $\alpha \in [-4.33, -1.5]$, with the errorbars indicating the 16th and 84th percentiles. The bottom panel shows the results across all $\alpha \in [-4.33, -1.5]$. These constraints are obtained by enforcing the observed location of the break in the $\tau_d$ distribution to be at $\sim 2$ Gyr. 
(Further details can be found in the text.) } 
\label{fig: Q0 histogram}
\end{figure}

\begin{figure*}[t!]
\centering
\includegraphics[width=\textwidth]{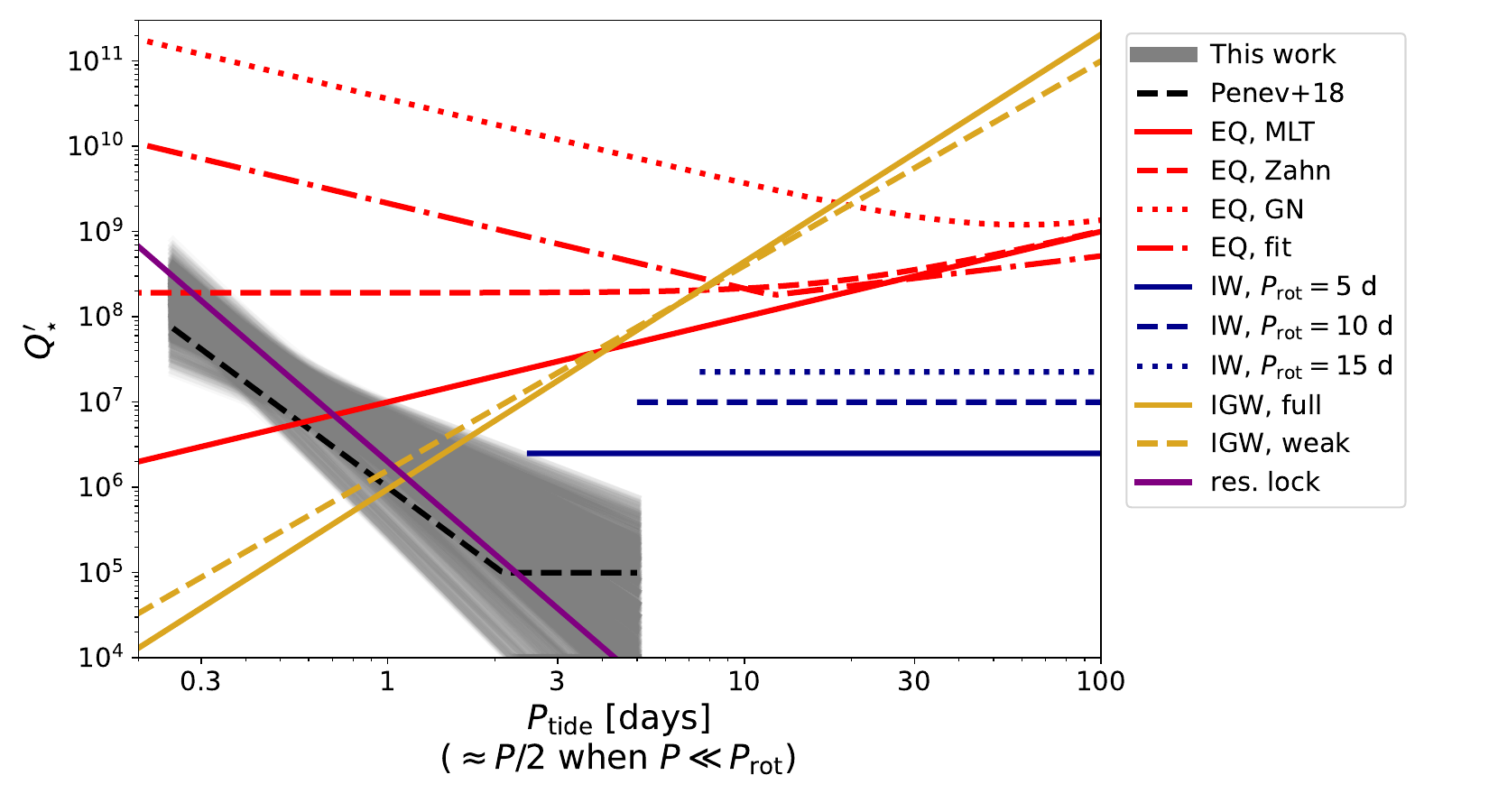}
\caption{Summary of empirical constraints on $Q_{\star}'$ vs. $P_{\mathrm{tide}}$. The gray lines represent constraints from the observed $\tau_d$ histograms of the AP sample limited to planets around cool stars only. Each line corresponds to a different combination of $\alpha \in [-4.33, -1.5]$ and $Q_0$ inferred from the data, as described in the text. The black dashed line shows the best-fitting saturated power law from \citetalias{2018AJ....155..165P}'s empirical constraints, which shows excellent agreement with our results. The red lines show the predictions from equilibrium tides with different prescriptions for the effective viscosity as described in the text: mixing length theory, \cite{1966AnAp...29..313Z}, \cite{1977Icar...30..301G}, and the best-fit numerical curve from \cite{2020MNRAS.497.3400D}. The blue lines show the frequency-averaged dissipation expected from inertial waves in convective zones, $Q_*' \approx 10^7 (P_{\rm rot}/10~{\rm d})^2$ \citep{2013MNRAS.429..613O,2020MNRAS.498.2270B} with different stellar rotation periods. The yellow lines show the predictions from internal gravity waves in radiative zones in the fully-damped wave breaking regime \citep{2020MNRAS.498.2270B} and the weakly nonlinear regime \citep{2016ApJ...816...18E}. The purple line shows the expectations from resonance locking based on \cite{2021ApJ...918...16M}. } 
\label{fig: Qstar vs P}
\end{figure*}

To this point we have focused on constraining $\alpha$, the exponent of $P$ in the two parameterizations of $Q_{\star}'$ (equations \ref{eq: Penev Q_star' with P approximation} and \ref{eq: Q_star' with stellar dynamical timescale}). We also need to constrain the other parameter, $Q_0$, which dictates the value of $Q_{\star}'$ when $P = 2$ days for the \citetalias{2018AJ....155..165P} parameterization or when $P = 100 t_{\star}$ for the $t_{\star}$ parameterization. So far we have fixed $Q_0 = 10^6$ without loss of generality, since the quantities $\alpha$ and $Q_0$ have different effects on the $\tau_d$ distribution. Whereas $\alpha$ changes the slope of the steady-state region, $Q_0$ maintains the overall shape of the $\tau_d$ distribution but shifts it systematically to lower or higher values. This is simply because $\tau_d \propto Q_0$.  

Here we use this shift effect on the $\tau_d$ distribution as a means of constraining $Q_0$. Our approach is to locate the break in the $\tau_d$ distribution when $Q_0 = 10^6$, which we will denote $\tau_{\mathrm{break},10^6}$. We then modify $Q_0$ until the observed break location lines up to theoretical expectation. Specifically, in Sections \ref{sec: toy model} and \ref{sec: analytic model}, the simulations and PDE solution showed that the break is located at $\tau_d \sim 1-3$ Gyr $\sim$ age. 
We will thus look for the value of $Q_0$ such that the break in the observed distribution of $\tau_d$ is at $\sim 2$ Gyr. Other choices are reasonable, and it straightforward to scale the constraints on $Q_0$ accordingly.

Let us consider the sample of all planets around cool stars and aim to derive holistic constraints on $Q_0$. We loop through the values of $\alpha \in [-4.33, -1.5]$, since this is the inferred range from Figure \ref{fig: slope_vs_alpha_all_planets_cool_vs_hot}. For each value, we first set $Q_0 = 10^6$ and use a bootstrap resampling method to generate 100 $\tau_d$ histograms. For each histogram, we fit a piecewise power law, letting the break location $\tau_{\mathrm{break},10^6}$ and the slopes of the two power laws be free parameters, and we infer the new value of $Q_0 = 10^6(2 \ \mathrm{Gyr}/\tau_{\mathrm{break},10^6})$. In this way, we obtain 100 values of $Q_0$ associated with each value of $\alpha$. In Figure \ref{fig: Q0 histogram}, we show the constraints on $Q_0 = 10^6(2 \ \mathrm{Gyr}/\tau_{\mathrm{break},10^6})$ for each $\alpha$ (top panel) and across all $\alpha \in [-4.33, -1.5]$ (bottom panel). Our results favor $Q_0 \in [10^{5.5}, 10^7]$. This checks out with expectation; for instance, the $\tau_d$ distribution with $\alpha = -3$ and $Q_0 = 10^6$ (Figure \ref{fig: observed distribution (all planets and hot jupiters) with alpha = -3}, top panel) shows a break at $\tau_{\mathrm{break},10^6}\sim 0.5 \ \mathrm{Gyr}$, and Figure \ref{fig: Q0 histogram} shows that $Q_0 \sim 10^{6.5}$.\footnote{For completeness, we also perform this procedure for the full sample of planets without any cut on $T_{\mathrm{eff}}$, and the results are nearly identical. We present the results for the cool star sample since we believe this sample is a more direct probe of tidal sculpting.}

\subsection{Empirical constraints on $Q_{\star}'$}
\label{sec: empirical constraints on Qstar'}

We can now summarize our constraints on both $Q_0$ and $\alpha$ by plotting the inferred $Q_{\star}'$ vs. $P_{\mathrm{tide}}$ in Figure \ref{fig: Qstar vs P}. Recall that we have been assuming $P_{\mathrm{tide}} \approx P/2$, but we use $P_{\mathrm{tide}}$ here to be more general. For each value of $\alpha \in [-4.33, -1.5]$, we sample from the 100 inferred values of $Q_0$ described in the previous paragraph, and we plot a gray line equal to $Q_{\star}' = Q_0 (P_{\mathrm{tide}}/ \mathrm{days})^{\alpha}$. These lines collectively indicate the region of $Q_{\star}'$ vs. $P_{\mathrm{tide}}$ space that is consistent with the data. The lines are only shown in the range $0.25 \ \mathrm{days} < P_{\mathrm{tide}} < 5 \ \mathrm{days}$ ($0.5 \ \mathrm{days} \lesssim P \lesssim 10 \ \mathrm{days}$), since this is the region where we have constraining power. Comparing our constraints with those of \citetalias{2018AJ....155..165P}, it is clear that we favor a similar dependence of $Q_{\star}'$ vs. $P_{\mathrm{tide}}$. We find a similar $\alpha$ and a similar but slightly larger $Q_0$, such that $Q_{\star}'$ is larger than \citetalias{2018AJ....155..165P}'s fit for the $P \sim 1-5$ days range where most hot Jupiters are found and where our constraints are the strongest. 

We can also compare our results to the empirical constraints from \cite{2018MNRAS.476.2542C}, who used a hierarchical Bayesian analysis to infer the average $Q_{\star}'$ based on the mass-separation distribution of hot Jupiters. They found $\log_{10}Q_{\star}' = 8.26 \pm 0.14$ (and $\log_{10}Q_{\star}' = 7.3 \pm 0.4$ for a subset of the sample expected to be more susceptible to dynamical tides). Although this is larger than our average constraints across all periods, if their results are dominated by the hot Jupiters with the very shortest periods, then they are compatible with our constraints. However, these comparisons may be complicated by other factors affecting the mass-separation distribution \citep[e.g.][]{2018MNRAS.479.5012O}.

Qualitatively, our results indicate a strongly frequency-dependent dissipation that is quite efficient at longer periods but weak at shorter periods. An inspiraling planet would thus spend more time at short periods than expected from a constant $Q_{\star}'$ evolution. Our constraints on $Q_{\star}'$ vs. $P_{\mathrm{tide}}$ offer the opportunity to compare with theoretical predictions of tidal dissipation models, which we will discuss in the next section.




\section{Discussion}
\label{sec: discussion}

\subsection{Theoretical models of tidal dissipation}
\label{sec: theoretical models}


Many theoretical models of tidal dissipation have been proposed over the years. The difficulty is in pinpointing the exact mechanisms by which tidal energy is dissipated in a fluid body. The models predict different amplitudes $Q_0$ and dependence on the tidal period $P_{\mathrm{tide}}$. 
The tidal response is typically decomposed into two components: an equilibrium tide and a dynamical tide \citep[e.g.][]{1966AnAp...29..313Z, 1975A&A....41..329Z, 1977A&A....57..383Z}; see  \cite{2014ARA&A..52..171O} for a review. The equilibrium tide represents the large-scale, quasi-hydrostatic deformation response to the tidal potential, which is thought to be dissipated by turbulent convection. The dynamical tide is a wave-like component that is excited by tidal forcing and dissipated by non-adiabatic processes and nonlinear effects. Below we will briefly review the various theoretical models and highlight their predictions of $Q_{\star}'$ vs. $P_{\mathrm{tide}}$. We will then holistically synthesize and compare these predictions with our constraints to assess which theories are favored. We largely follow \cite{2020MNRAS.498.2270B}, who recently reviewed these dissipation mechanisms with a focus on star-planet tidal interactions.

\newpage
\subsubsection{Equilibrium tides}
\label{sec: equilibrium tides}
In terms of the equilibrium tide, the most commonly-used approximation has been weak friction model which adopts a constant time lag of the tidal response at all tidal frequencies \citep[e.g.][]{1880RSPT..171..713D, 1973Ap&SS..23..459A, 1981A&A....99..126H}. However, there is no a priori reason to expect a lack of dependence on the tidal frequency, so various models have relaxed this assumption. The interaction between the equilibrium tide and turbulent convection is often thought to act like an effective viscosity $\nu_E$ that damps the large-scale flow. Mixing-length theory (MLT) predicts a frequency-independent $\nu_E$. However, the dissipation is expected to be less efficient when the tidal frequency $\omega \equiv 2\pi/P_{\mathrm{tide}}$ exceeds the turnover frequency of the dominant, largest-scale convective eddies $\omega_c \sim  v_c/L_c$, where $v_c$ is the convective velocity and $L_c \sim H$ is the convective length scale, which is related to the scale height $H$ in MLT. When the tidal frequency is higher, tides couple less effectively to the slower turbulently convective motions \citep[a recent comparison of the possible physical explanations is provided by][]{2020MNRAS.497.3400D}.  In particular, \cite{1966AnAp...29..313Z} proposed a form $\nu_E \propto (\omega_c/\omega)$, whereas \cite{1977Icar...30..301G} suggested $\nu_E \propto (\omega_c/\omega)^2$. The quadratic scaling is favored at high frequencies by recent hydrodynamical simulations of tidal flows, such as \cite{2020MNRAS.497.3400D} and \cite{2020MNRAS.497.4472V}. \cite{2020MNRAS.497.3400D} used their simulations to obtain a continuous power-law fit of $\nu_E$ as a function of $\omega/\omega_c$. Figure \ref{fig: Qstar vs P} shows four curves of $Q_{\star}'$ vs. $P_{\mathrm{tide}}$ corresponding to these four different effective viscosity prescriptions. They are labeled ``MLT'', ``Zahn'', ``GN'' and ``fit''. 

\subsubsection{Dynamical tides}
\label{sec: dynamical tides}
Dynamical tides can theoretically propagate in any region of the star that supports waves. The waves most effectively driven by tides tend to be those with similar frequencies to the tidal forcing. Thus, while high-frequency acoustic modes propagate in all stellar regions, they often don't factor substantially in tidal dissipation. On the other hand, convective zones support inertial waves, which are excited by tidal forcing and restored by the Coriolis force \citep[e.g.][]{2007ApJ...661.1180O, 2009ApJ...696.2054G, 2009MNRAS.396..794O, 2010MNRAS.407.1631P, 2013MNRAS.429..613O, 2015A&A...580L...3M, 2017A&A...604A.112G, 2022ApJ...927L..36B}. Inertial waves have a complicated frequency dependence due to a forest of resonances in the range $P_{\mathrm{tide}} > P_{\mathrm{rot}}/2$, so it is common to compute a frequency-averaged dissipation across this interval \citep[e.g.][]{2020MNRAS.498.2270B}. Figure \ref{fig: Qstar vs P} shows three lines representing the expected magnitude of inertial waves for different stellar rotation periods based on the results of \cite{2013MNRAS.429..613O}. The mechanism only operates for tidal periods longer than half the rotation period of the star. Inertial waves are thus probably not as relevant for short-period planets around main sequence stars, but they may be much more influential during the pre-main sequence phase. 

In contrast to inertial waves in convective zones, radiative zones support internal gravity waves ($g$-modes), which are restored by buoyancy \citep{1998ApJ...507..938G, 2007ApJ...661.1180O, 2010MNRAS.404.1849B, 2012ApJ...751..136W, 2016ApJ...816...18E, 2020MNRAS.495.1239S, 2024ApJ...960...50W}. A common assumption is that the gravity waves are excited by tidal forcing at the interface between the convective and radiative zones of the star and propagate inwards, where they are damped in the radiative zone. As the waves travel inwards, they become geometrically focused and their amplitudes increase, leading the waves to become nonlinear and potentially even break. A fully damped, wave-breaking scenario requires the companion mass to exceed a critical threshold which depends on the structure of the star but is approximately $3.3 \ M_{\mathrm{Jup}} (P/{1 \ \mathrm{day}})^{-1/6}$ for the current Sun \citep{2020MNRAS.498.2270B}. If breaking occurs, the internal gravity waves are traveling waves and dissipation is expected to be very efficient, scaling as $Q_{\star}' \approx  1.5 \times 10^5 (P_{\mathrm{tide}}/0.5 \ \mathrm{d})^{8/3}$. \cite{2016ApJ...816...18E} and \cite{2024ApJ...960...50W} studied a weakly nonlinear regime where the primary waves do not break but are sufficiently nonlinear to excite secondary waves that dissipate their energy. This weakly nonlinear regime creates standing waves and still predicts efficient damping with a slightly less steep slope, $Q_{\star}' \approx 3 \times 10^5 (M_p/M_{\mathrm{Jup}})^{0.5} (P_{\mathrm{tide}}/0.5 \ \mathrm{d})^{2.4}$ \citep{2016ApJ...816...18E}. Figure \ref{fig: Qstar vs P} shows the predictions from both of these regimes in yellow.

A variation of internal gravity wave dissipation is a mechanism called ``resonance locking'' \citep{1999A&A...350..129W, 2001A&A...366..840W, 2012MNRAS.420.3126F, 2013MNRAS.433..332B, 2017MNRAS.472.1538F, 2021AJ....161..263Z, 2021ApJ...918...16M, 2024ApJ...967L..29Z}. Gravity modes in the radiative zone have a dense frequency spectrum. Resonance locking occurs when a planet becomes trapped in a resonance between the tidal forcing frequency and a $g$-mode oscillation frequency of the star. As the star evolves and its internal structure changes, the frequencies change but the resonance is maintained. Resonance locking yields efficient tidal dissipation and fast migration that is typically inward for a planet orbiting a star \citep{2021ApJ...918...16M}. The planet's orbit decays on a mode evolution timescale that is nearly independent of the planet's orbital period, such that the effective tidal quality factor is larger for shorter orbital periods. Using MESA models to compute oscillation modes, \cite{2021ApJ...918...16M} made predictions for the effective stellar quality factor due to resonance locking. An approximation to their result is shown in Figure \ref{fig: Qstar vs P}. However, \cite{2021ApJ...918...16M} noted a theoretical tension with this prediction that will be discussed below.

\subsubsection{Comparison with empirical constraints}
\label{sec: synthesis}

We can now consider the theoretical predictions of $Q_{\star}'$ altogether and compare them to our constraints. Equilibrium tides appear to be too inefficient, and the slopes are not steep enough. The best-fit numerical curve from \cite{2020MNRAS.497.3400D} does not overlap with the constraints. Inertial waves are fairly efficient, but they are only relevant to the range of $P_{\mathrm{tide}} > P_{\mathrm{rot}}/2$, which corresponds to longer tidal periods than the regime we are discussing. 

Internal gravity waves produce highly efficient dissipation at short tidal periods, but the predicted slope is in the opposite direction for both the wave breaking and weakly nonlinear regimes. Assuming our constraints are correct, it does not appear that these regimes dominate. Most planets are not above the critical mass for wave breaking \citep{2020MNRAS.498.2270B}, so on the one hand it is reasonable that the predictions for this case do not fit. On the other hand, \cite{2023MNRAS.521.1353G} recently showed using hydrodynamical simulations that fully damped internal gravity waves may operate for a significantly wider regime of planetary masses due to the waves transferring angular momentum to the central region of the star and generating an expanding critical layer that absorbs subsequent waves and causes them to break. Further work is necessary to investigate why neither this fully damped regime nor the weakly nonlinear regime appear to fit our constraints. We will also discuss this further below.

Taken at face value, \cite{2021ApJ...918...16M}'s predictions from resonance locking are by far the best fit to the empirical constraints from this work and from \cite{2018AJ....155..165P}. This is somewhat surprising for reasons discussed in \cite{2021ApJ...918...16M}. Specifically, resonance locking is based on a linear analysis of dynamical tides.  \cite{2021ApJ...918...16M} argued that the weakly nonlinear damping of $g$-modes proposed by \cite{2016ApJ...816...18E} would saturate resonant mode excitation and prevent resonance locking from occurring for hot Jupiters around Sun-like stars. Thus, \cite{2021ApJ...918...16M} urged caution to the interpretation of their $Q_{\star}'$ vs. $P_{\mathrm{tide}}$ prediction from resonance locking.

The obstacle of nonlinear damping  was recently revisited by \cite{2024ApJ...967L..29Z}, who proposed a solution based on the idea that the secondary modes are strong enough to break and become traveling waves rather than standing waves. If this is correct, they argued that resonance locking is effective out to $a/R_{\star} \approx 8-10$ ($P_{\rm orb} \lesssim 4$~days), which would be consistent with the constraints from this work. However, this picture has not yet been verified by hydrodynamical simulations. Recently, \cite{2023MNRAS.521.1353G} performed hydrodynamical simulations of tidally-excited gravity waves focusing on $m=2$ modes and found that resonance locking conditions can be significantly altered by a differentially rotating core driven by the nonlinear feedback of the waves. Although \cite{2024ApJ...967L..29Z} primarily invoked axisymmetric $m=0$ modes, which would likely not produce the same differential rotation in the stellar core as the $m=2$ waves, the scenario with $m=0$ modes should be investigated with detailed simulations.

While further work is needed, it is also worth emphasizing that  \cite{2024ApJ...967L..29Z} argued that resonance locking offers a compelling explanation for the stellar obliquity trends with effective temperature mentioned in Section \ref{sec: hot/cool stars}. Hot Jupiters around cooler stars with radiative cores simultaneously experience spin-orbit realignment and orbital decay, while hot Jupiters around hot stars that lack radiative cores do not experience efficient tidal evolution.  In this work we posited that the tidal decay timescale distribution shows evidence of tidal evolution for planets around cool stars but not hot stars. This would also be consistent with a resonance locking scenario, since the $g$-mode frequencies of hot stars do not change much during their lifetimes \citep{2024ApJ...967L..29Z}.

Taken together, our constraints on $Q_{\star}'$ agree most with the predictions of resonance locking among the currently-discussed dissipation mechanisms. It is important to note, however, that there may not necessarily be a single dominant mechanism of tidal dissipation. It is possible that multiple mechanisms operate in main sequence stars, and it is even more likely that the primary mechanism changes for different stellar types and ages. For instance, during the pre-main sequence phase, inertial waves are more important due to the rapid rotation of the host star \citep[e.g.][]{2020MNRAS.498.2270B, 2022ApJ...927...22S}.  Such a combination of dissipation mechanisms could also produce the observed constraints on $Q_{\star}'$.   It is beyond the scope of this work to study the impact of these considerations on the tidal decay timescale distribution, but this would be beneficial in future work.




\subsection{Comparison to observed planetary systems}
\label{sec: comparison to observed}

A few planets have been observed to be spiraling inwards in real time. Here we investigate how their measured dissipation rates compare to our population-level constraints. WASP-12 b\footnote{WASP-12 b was not included in our observational sample in Section \ref{sec: observations} because its measured eccentricity is $0.0317 \pm 0.0087$ \citep{2020ApJ...888L...5Y}, higher than our 0.02 cutoff.} was the first planet confirmed to be decaying \citep{2016A&A...588L...6M, 2017AJ....154....4P, 2020ApJ...888L...5Y, 2021AJ....161...72T}. The measurement by \cite{2024A&A...685A..63A} yielded $P/{\dot{P}} = -3.13 \pm 0.087$ Myr. If the decay is due to stellar tides, this corresponds to $Q_{\star}' = 1.70 \pm 0.14 \times 10^5$. WASP-12 b has an orbital period of 1.09 days, and the rotation period of the star is $\sim 36$ days \citep{2010MNRAS.405.2037W, 2019MNRAS.482.1872B}. This places the inferred $Q_{\star}'$ much below our average empirical constraints at this period ($Q_{\star}' \approx 10^{6.5}(P_{\mathrm{tide}}/\mathrm{days})^{-3} \approx 1.8\times10^7$, Figure \ref{fig: Qstar vs P}). 

There are a couple ways to interpret this. First, it's possible that WASP-12 b's orbital decay is dominated by tides in the planet rather than in the star (\citealt{2018ApJ...869L..15M}, though see \citealt{2022MNRAS.509.3301S}). Second, the host star may have evolved onto the subgiant branch and lost its convective core, which would allow for the possibility that the tidal dissipation is very efficient due to nonlinear wave-breaking of internal gravity waves near the star's center \citep{2017ApJ...849L..11W, 2019MNRAS.482.1872B}. However, there are some caveats to this interpretation \citep{2019MNRAS.482.1872B}. Alternatively, there is evidence for another relevant tidal mechanism for main sequence F stars, in which internal gravity waves are converted to outwardly-propagating magnetic waves, which become fully damped as they travel through the radiative zone \citep{2024ApJ...966L..14D}. This extends the fully damped regime for internal gravity waves.

We stress that we do not expect every observed system to fit our population-level constraints. To the extent that our constraints are an accurate reflection of tidal dissipation efficiency across the population, we are probing the dominant mechanisms and average interior structures for main sequence stars. In this context, we expect outliers because even main sequence stars have distinct interiors with varying mass \citep{2012ApJ...757....6H}. As stars evolve, even more dramatic changes occur, and dissipation is often more efficient during the pre-main sequence and subgiant phases when convective zones occupy a larger fraction of the stellar mass \citep[e.g.][]{2020MNRAS.498.2270B}. We note that the most detectable cases of orbital decay might preferentially highlight outliers toward rapid decay. 

\begin{figure}[t!]
\centering
\includegraphics[width=\columnwidth]{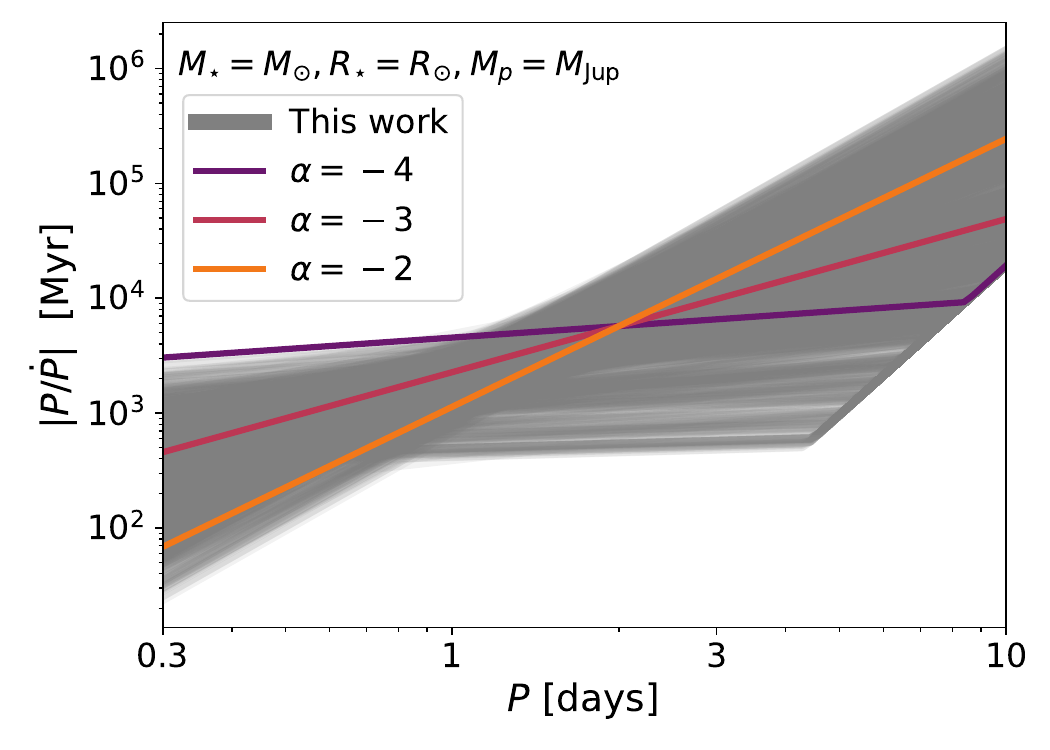}
\caption{$|P/\dot{P}|$ versus orbital period for a Jupiter-mass planet orbiting a Solar-like star. The gray lines represent our empirical constraints as in Figure \ref{fig: Qstar vs P}. The other lines represent equation \ref{eq: P/Pdot for different alpha} and use fiducial values of $\alpha$ and $Q_0 = 10^{6.5}$. A value of $\alpha = -4.33$ corresponds to a flat line. The cutoff on the right hand side is a result of hitting $\min(Q_{\star}') = 10^4$.} 
\label{fig: P/Pdot vs P}
\end{figure}

The Kepler-1658 (KOI-4) system tells a slightly different story. The host star is evolved ($1.5 \ M_{\odot}$, $2.9 \ R_{\odot}$). \cite{2022ApJ...941L..31V} detected a decreasing orbital period $\dot{P} = -131^{+20}_{-22} \ \mathrm{ms \ yr^{-1}}$, corresponding to $P/\dot{P} \approx -2.5$ Myr and a stellar quality factor equal to $Q_{\star}' = 2.50^{+0.85}_{-0.62} \times 10^4$. The orbital period is 3.85 days, and the stellar rotation period is 5.66 days, such that $P_{\mathrm{tide}} \approx 6.02$ days. Our average empirical constraints at this period yield $Q_{\star}' \approx 10^{6.5}(P_{\mathrm{tide}}/\mathrm{days})^{-3} \approx 1.4\times10^4$, approximately consistent with \cite{2022ApJ...941L..31V}'s derived constraints. This suggests that perhaps resonance locking is responsible for Kepler-1658's orbital decay. An alternative possibility is inertial waves in the convective envelope during the subgiant phase, although \cite{2024MNRAS.527.5131B} show that this period of evolution is very short-lived ($\lesssim 100$ years).

Searches for orbital decay across the hot Jupiter population have returned no additional confident detections beyond the two aforementioned systems \citep[e.g.][]{2022ApJS..259...62I, 2024arXiv240407339A}. \cite{2024arXiv240407339A} studied a sample of 43 hot Jupiters with updated timing information and showed that nearly half of the targets must not be experiencing orbital decay as rapid as WASP-12 b, since they would have been detected otherwise. For some of the systems, they inferred that the $Q_{\star}'$ must be at least an order of magnitude greater than that of WASP-12. This result is qualitatively consistent with our empirical constraints. Namely, at the short-period end, we find evidence for fairly large $Q_{\star}' \approx  10^7-10^8$ for $P \approx 1$ day (Figure \ref{fig: Qstar vs P}). This suggests that detectable orbital decay may be rare, a prediction that can be tested with continued timing monitoring.

To aid future comparison to observations, we close this section with a few relations. Note that 
\begin{equation}
\frac{P}{\dot{P}} = \frac{2}{3}\tau_a = -\frac{13+3\alpha}{3}\tau_d.
\end{equation}
Using equation \ref{eq: tau_a Penev version} with arbitrary $\alpha$,
\begin{equation}
\frac{P}{\dot{P}} = \frac{2 Q_0}{3 C}\left(\frac{M_{\star}}{M_{\odot}}\right)^{\frac{1-\alpha}{2}}\left(\frac{R_{\star}}{R_{\odot}}\right)^{-5}\left(\frac{M_p}{M_{\mathrm{Jup}}}\right)^{-1}\left(\frac{a}{\mathrm{AU}}\right)^{\frac{13+3\alpha}{2}}
\end{equation}
where $C$ is in equation \ref{eq: C Penev version}.
For various values of $\alpha$, 
\begin{equation}
\begin{split}
\alpha = -4: \ \ \frac{P}{\dot{P}} &= -5625 \ \mathrm{Myr}\left(\frac{Q_0}{10^{6.5}}\right)\left(\frac{M_{\star}}{M_{\odot}}\right)^{\frac{5}{2}} \\ 
& \times \left(\frac{R_{\star}}{R_{\odot}}\right)^{-5}\left(\frac{M_p}{M_{\mathrm{Jup}}}\right)^{-1}\left(\frac{a}{0.03 \ \mathrm{AU}}\right)^{\frac{1}{2}} \\
\alpha = -3: \ \ \frac{P}{\dot{P}} &= -5337 \ \mathrm{Myr}\left(\frac{Q_0}{10^{6.5}}\right)\left(\frac{M_{\star}}{M_{\odot}}\right)^{2} \\ 
& \times \left(\frac{R_{\star}}{R_{\odot}}\right)^{-5}\left(\frac{M_p}{M_{\mathrm{Jup}}}\right)^{-1}\left(\frac{a}{0.03 \ \mathrm{AU}}\right)^{2} \\
\alpha = -2: \ \ \frac{P}{\dot{P}} &= -5063 \ \mathrm{Myr}\left(\frac{Q_0}{10^{6.5}}\right)\left(\frac{M_{\star}}{M_{\odot}}\right)^{\frac{3}{2}} \\ 
& \times \left(\frac{R_{\star}}{R_{\odot}}\right)^{-5}\left(\frac{M_p}{M_{\mathrm{Jup}}}\right)^{-1}\left(\frac{a}{0.03 \ \mathrm{AU}}\right)^{\frac{7}{2}} \\
\alpha = 0: \ \ \frac{P}{\dot{P}} &= -4558 \ \mathrm{Myr}\left(\frac{Q_0}{10^{6.5}}\right)\left(\frac{M_{\star}}{M_{\odot}}\right)^{\frac{1}{2}} \\ 
& \times \left(\frac{R_{\star}}{R_{\odot}}\right)^{-5}\left(\frac{M_p}{M_{\mathrm{Jup}}}\right)^{-1}\left(\frac{a}{0.03 \ \mathrm{AU}}\right)^{\frac{13}{2}}.
\label{eq: P/Pdot for different alpha}
\end{split}
\end{equation}
These relations can be used to quickly estimate the predicted orbital decay rate given different assumptions about $\alpha$ and $Q_0$. Figure \ref{fig: P/Pdot vs P} summarizes these relations and our empirical constraints on $P/{\dot{P}}$ for the case of a Jupiter-mass planet orbiting a Solar-like star. 

\subsection{Caveats and extensions of the model}
\label{sec: caveats and extensions}

The empirical framework we developed in this paper is the first of its kind, and we made a number of assumptions for the sake of simplicity and tractability of the model. However, it is possible that some of these assumptions may impact our results at some level. Here we outline a few caveats and note how the model could be improved in the future.

In our synthetic hot Jupiter simulations, we assumed that stellar tides were the only source of orbital evolution and that there were no planetary tides (i.e. zero eccentricity and planetary obliquity). Then, when working with the observed planet population, we only considered planets with constraints consistent with circular orbits. However, planets may have been eccentric in the past. Indeed, if high eccentricity migration is a common mechanism for hot Jupiter formation \citep{2024Natur.632...50G}, then planetary tides would have played a significant role in the past orbital evolution. This history could be indirectly accounted for by just defining the initial semi-major axis in the synthetic population model as that which a planet attains after undergoing high eccentricity migration and ending on a circular orbit, as noted in Section \ref{sec: toy model}. However, future updates to the model could try to incorporate planetary tides directly.

Similar to eccentricities, our model has not accounted for the evolution of stellar rotations or spin-orbit misalignments. This is largely a practical concern, since observational constraints on these parameters are not available for all systems. It is not immediately clear how much these considerations could affect our results because it depends on the tidal mechanisms at play. One option for future updates would be to adopt theoretical models of stellar rotational and obliquity evolution and explore how this affects the decay timescale distribution.

Another assumption we made in our model was a simple, time-independent parameterization of $Q_{\star}'$ that only depended on the planet's orbital period. In reality, there may be certain points in time where the dominant tidal mechanism changes abruptly, particularly during stellar evolution transitions from pre-main sequence to main sequence and post-main sequence. If a tidal mechanism like wave breaking suddenly turns on or off \citep[e.g.][]{2020MNRAS.498.2270B}, then the effective $Q_{\star}'$ would also change suddenly. Modeling such variations was beyond the scope of this work but could be considered in the future.

\section{Conclusions}
\label{sec: conclusions}

Numerous past studies have analyzed signatures of tidal evolution in the short-period exoplanet population, but it has remained challenging to determine the mechanisms and associated efficiencies of the dissipation of stellar tides. We introduced a novel population-level technique based on the tidal decay timescale $\tau_d$ to constrain the stellar tidal quality factor $Q_{\star}'$. Code reproducing some of our results can be found at \url{https://github.com/smillholland/Hot_Jupiter_Tides}. Our main takeaways are as follows:
\begin{enumerate}
\item When expressed in terms of the tidal decay timescale, the distribution of exoplanets satisfies a continuity equation and exhibits a steady-state region for $\tau_d$ less than the mean system age  (Figures \ref{fig: tau_decay distribution}--\ref{fig: PDE solution}). 

\item  The properties of the steady-state region probe the tidal dissipation efficiency, $Q_{\star}'$.  
We used a sample of observed planets to evaluate the power law exponent $\alpha$ and normalization $Q_0$ of a period-dependent parameterization of $Q_{\star}'$ (equation \ref{eq: Penev Q_star' with P approximation} and Section \ref{sec: observations}).
The population is best described by a steep period dependence indicating less efficient dissipation at short orbital periods, with $\alpha \in [-4.33, -2]$ and $Q_0 \in [10^{5.5}, 10^7]$ (Figures \ref{fig: slope_vs_alpha}, \ref{fig: slope_vs_alpha_all_planets_cool_vs_hot}, \ref{fig: Qstar vs P}).

\item While  cool star systems show clear evidence of tidal sculpting, hot star systems do not (Figure \ref{fig: slope_vs_alpha_all_planets_cool_vs_hot}). This result may be consistent with expectations for different tidal dissipation rates for these populations.

\item We compared our constraints on $Q_{\star}'$ vs. $P_{\mathrm{tide}}$ to  the theorized mechanisms of tidal dissipation in Section \ref{sec: theoretical models} and Figure \ref{fig: Qstar vs P}. Equilibrium tides, inertial waves, and internal gravity waves do not produce a good fit to the constraints, at least not when acting in isolation. Resonance locking predictions \citep{2021ApJ...918...16M} provide an excellent fit. If this strong agreement indicates that resonance locking is indeed operating, this would imply that it is not suppressed by nonlinear damping \citep{2024ApJ...967L..29Z}.

\item There are two planetary systems with measured orbital decay. Of these,   WASP-12 b is decaying more rapidly than predicted by the population model (lower $Q_{\star}'$), while Kepler-1658 b matches the population-level prediction. We suggest that this individual dispersion might be expected  due to  variations in stellar structures and dissipation mechanisms at play.

\item The $|P/\dot{P}|$ timescales predicted by our population model are given in equation \ref{eq: P/Pdot for different alpha} and Figure \ref{fig: P/Pdot vs P}. Our results suggest that detectable orbital decay is rare, a prediction which is consistent with transit timing monitoring so far \citep{2022ApJS..259...62I, 2024arXiv240407339A} and which will continue to be tested with further monitoring.

\item The tidal disruption  of exoplanets (equation \ref{eq: disruption rate}) is the end point of tidal decay. Future observations of planet disruptions might offer a parallel constraint on tidal dissipation efficiency and the supply of planets into decaying orbits.

\end{enumerate}

Our work offers the first analysis that exploits the distribution of tidal decay timescales in order to understand the population as the solution to a continuity equation. We explored some divisions of the population (e.g. hot vs. cool stars), but the analysis could be productively extended to further subdivisions (e.g. planet types, spectral types, main sequence vs. subgiants) to test the universality of tidal mechanisms for different types of planets and host stars. It would also be useful to incorporate more information from theoretical stellar models \cite[e.g.][]{2012ApJ...757....6H} as an alternative to the empirical approach we adopted here.

\section{Acknowledgements}

We are grateful to the reviewer, Adrian Barker, for his thorough and constructive comments, which we believe have greatly improved the paper. We thank Jim Fuller, Linhao Ma, Kaloyan Penev, Yubo Su, Josh Winn, and J.~J. Zanazzi for valuable comments on a draft of this manuscript. In addition, we thank Kishalay De, Max Goldberg, Veome Kapil, Abraham Loeb, and Michelle Vick for helpful conversations about this and related topics. MM gratefully acknowledges support from the Clay Postdoctoral Fellowship of the Smithsonian Astrophysical Observatory. This research has made use of the NASA Exoplanet Archive, which is operated by the California Institute of Technology, under contract with the National Aeronautics and Space Administration under the Exoplanet Exploration Program.

\appendix 

\section{Alternative derivation of decay timescale distribution}
\label{sec: alternative derivation}

In Section \ref{sec: analytic model}, we derived the $\tau_d$ distribution using a continuity equation. Here we present an alternative derivation considering only the evolution of planets within bins in $\tau_d$ space, which is more directly analogous to our explorations using a simulated planet population in Section \ref{sec: toy model}. We will show that it yields an identical PDE to equation \ref{eq: continuity equation for x = tau}. For convenience, here we will shorten $\tau_d$ to simply $\tau$.

Consider the set of $\Delta N_i$ planets with $\tau$ in the small interval [$\tau_i$, $\tau_{i+1}$] at time $t_j$. Let $\rho(\tau, t)$ be the distribution such that 
\begin{equation}
\Delta N_i = \rho(\tau_i, t_j)\Delta \tau
\label{eq: N_i}
\end{equation}
where $\Delta \tau \equiv \tau_{i+1}-\tau_i$. Moreover, let $S(\tau)$ be the source distribution such that 
\begin{equation}
\Delta N_{\mathrm{born},i} = S(\tau_i)\Delta\tau\Delta t
\label{eq: N_born}
\end{equation} 
is the number of planets born in the $\Delta\tau$ range in a small window of time, $\Delta t \equiv t_{j+1}-t_j$.

Within the time $\Delta t$, the change in the number of planets in the $\tau$ bin is equal to 
\begin{equation}
\rho(\tau_i,t_{j+1})\Delta\tau - \rho(\tau_i,t_j)\Delta\tau = S(\tau_i)\Delta \tau \Delta t + \rho(\tau_{i+1},t_j) \Delta t - \rho(\tau_{i},t_j) \Delta t.
\label{eq: Delta relation 1}
\end{equation}
The first term on the right corresponds to the number of planets ``born'' into the $\tau$ bin in time $\Delta t$. The second term corresponds to the number of planets decaying into the bin from larger values of $\tau$. The third term corresponds to the number of planets decaying out of the bin to smaller values of $\tau$. Dividing through by $\Delta t \Delta\tau$, we have
\begin{equation}
\frac{\rho(\tau_i,t_{j+1}) - \rho(\tau_i,t_j)}{\Delta t} = S(\tau_i) + \frac{\rho(\tau_{i+1},t_j) - \rho(\tau_{i}, t_j)}{\Delta\tau}.
\label{eq: Delta relation 2}
\end{equation}
Taking the limit of small $\Delta t$ and $\Delta \tau$, we obtain
\begin{equation}
\frac{\partial\rho}{\partial t}\Bigr\rvert_{(\tau_i,t_j)} = S(\tau_i) + \frac{\partial\rho}{\partial\tau}\Bigr\rvert_{(\tau_i,t_j)}.
\label{eq: Delta relation 2}
\end{equation}
Finally, generalizing this to all $\tau$ and all $t$, we arrive at
\begin{equation}
\frac{\partial\rho}{\partial t} - \frac{\partial \rho}{\partial\tau} = S(\tau),
\label{eq: continuity equation for tau}
\end{equation}
which is identical to our result from Section \ref{sec: analytic model} (equation \ref{eq: continuity equation for x = tau} with $x = \tau_d$).

\section{Sensitivity to the source distribution}
\label{sec: source distribution}

The slope of the steady-state region is sensitive to the source distribution, particularly the assumed initial semi-major axis distribution. Here we define the initial semi-major axis to be that which a planet attains either after \textit{in situ} formation or after the initial migration (via disk torques or high-eccentricity tidal migration) has finished. The initial semi-major axis distribution of the true planet population is not well-known. Here we develop four variations of the simulated distributions of hot Jupiter populations from Section \ref{sec: toy model} using different choices for the initial semi-major axis distribution. We quantify the sensitivity of the slope of the steady-state region to these changes. 

We consider four initial semi-major axis distributions:
\begin{enumerate}
\item \textit{Uniform}: We used a uniform distribution in Section \ref{sec: toy model} and elsewhere throughout the main text. The probability density function (PDF) is
\begin{equation}
f(a; a_{\min}, a_{\max}) = \frac{1}{a_{\max} - a_{\min}},
\end{equation}
and we take $a_{\min} = 0.01$ AU and $a_{\max} = 0.1$ AU. 

\item \textit{Normal}: We consider a normal distribution that closely approximates the semi-major axis distribution of observed hot Jupiters. The PDF is
\begin{equation}
f(a; \mu, \sigma) = \frac{1}{\sigma \sqrt{2\pi}} \exp\left(-\frac{(a - \mu)^2}{2\sigma^2}\right)
\label{eq: normal}
\end{equation}
with $\mu = 0.04$ AU and $\sigma = 0.015$ AU. We truncate the distribution at $a_{\min}$ and $a_{\max}$ with values noted above.

\item \textit{Log-normal}: Next we consider a log-normal distribution which gives more probability at larger semi-major axes relative to the normal distribution. The PDF is
\begin{equation}
f(a; \mu, \sigma) = \frac{1}{a \sigma \sqrt{2\pi} \ln(10)} \exp\left(-\frac{(\log_{10}(a) - \mu)^2}{2\sigma^2}\right)
\label{eq: log-normal}
\end{equation}
with $\mu = -1.25$ and $\sigma = 0.2$. We again truncate the distribution at $a_{\min}$ and $a_{\max}$.

\item \textit{Power law}: Lastly, we consider a power law distribution ranging from $a_{\min}$ to $a_{\max}$. The PDF is
\begin{equation}
f(a; \beta, a_{\min}, a_{\max}) = \frac{\beta + 1}{a_{\max}^{\beta + 1} - a_{\min}^{\beta + 1}} a^\beta,
\label{eq: power law}
\end{equation}
and we take $\beta = 0.75$, which is motivated by the occurrence rate distribution of short-period giant planets \citep{2018AJ....155...89P}.
\end{enumerate}

\begin{figure*}[t!]
\centering
\includegraphics[width=\textwidth]{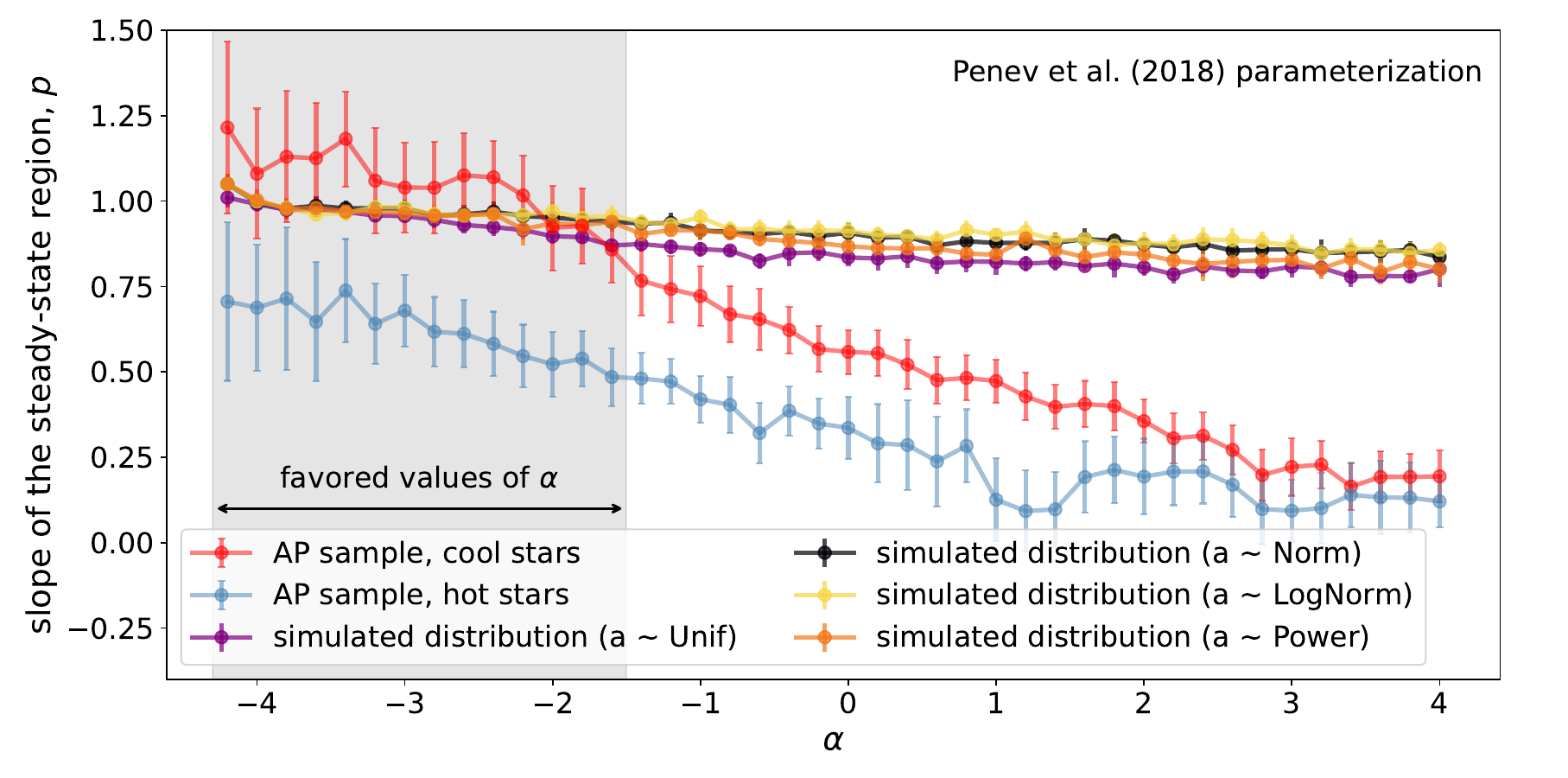}
\caption{Same as Figure \ref{fig: slope_vs_alpha_all_planets_cool_vs_hot} but we now show the sensitivity of the simulated curves to the initial semi-major axis distribution. The purple curve is the same as Figures \ref{fig: slope_vs_alpha} and \ref{fig: slope_vs_alpha_all_planets_cool_vs_hot} and corresponds to a uniform distribution of initial semi-major axes. The black, yellow, and orange curves correspond to the normal, log-normal, and power law distributions, respectively (equations \ref{eq: normal} through \ref{eq: power law}).} 
\label{fig: slope_vs_alpha_sensitivity_to_source_distribution}
\end{figure*}

For each of these initial semi-major axis distributions, we repeat the analysis of Section \ref{sec: results using P18 parameterization}, calculating the slope of the steady-state region as a function of $\alpha$. Figure \ref{fig: slope_vs_alpha_sensitivity_to_source_distribution} shows the analog of Figure \ref{fig: slope_vs_alpha_all_planets_cool_vs_hot} but now with multiple lines indicating the results from the different simulated distributions and with the analytic method from the PDE solution left out. We observe some variations in the slope calculations for the four curves. However, the deviations are small overall and have no significant implications for our findings.

\bibliographystyle{aasjournal}
\bibliography{main}

\end{document}